\documentclass[twocolumn,showpacs,preprintnumbers,nofootinbib,amsmath,aps,pra,longbibliography,amssymb,superscriptaddress]{revtex4-1}

\usepackage[export]{adjustbox}
\usepackage{amsfonts}
\usepackage{dsfont}
\usepackage{amsmath}
\usepackage[makeroom]{cancel}
\usepackage[english]{babel}
\usepackage[T1]{fontenc}
\usepackage{txfontsb}
\usepackage{times}
\usepackage{mathrsfs}
\usepackage{graphicx}
\usepackage{dcolumn}
\usepackage{bm}
\usepackage{wasysym}
\usepackage[colorlinks,bookmarks=true,citecolor=blue,linkcolor=red,urlcolor=blue]{hyperref}
\usepackage[tight, FIGTOPCAP, hang, raggedright, nooneline]{subfigure}
\usepackage{hyperref}
\usepackage[table]{xcolor}
\usepackage{epsfig}
\usepackage{amssymb}
\usepackage{lipsum}
\usepackage{enumitem}

\usepackage{qcircuit}

\makeatletter
\def\p@subsection{}
\makeatother

\newcommand{\subfigimg}[3][,]{%
\setbox1=\hbox{\includegraphics[#1]{#3}}
\leavevmode\rlap{\usebox1}
\rlap{\hspace*{-3pt}\raisebox{\dimexpr\ht1-.1\baselineskip}{#2}}
\phantom{\usebox1}
}

\definecolor{Green}{RGB}{80,182,0}

\usepackage{color}

\newcommand{\la}{\langle}
\newcommand{\ra}{\rangle}

\graphicspath{./Images/}
\usepackage{epstopdf}

\begin{document}
\title{Simulating quantum many-body dynamics on a current digital quantum computer}
\author{Adam Smith}
\email{adam.smith@tum.de}
\affiliation{Blackett Laboratory, Imperial College London, London SW7 2AZ, United Kingdom}
\affiliation{Department of Physics, T42, Technische Universit{\"a}t M{\"u}nchen, James-Franck-Stra{\ss}e 1, D-85748 Garching, Germany}
\author{M.~S.~Kim}
\affiliation{Blackett Laboratory, Imperial College London, London SW7 2AZ, United Kingdom}
\author{Frank Pollmann}
\affiliation{Department of Physics, T42, Technische Universit{\"a}t M{\"u}nchen, James-Franck-Stra{\ss}e 1, D-85748 Garching, Germany}
\author{Johannes Knolle}
\affiliation{Blackett Laboratory, Imperial College London, London SW7 2AZ, United Kingdom}
\date{\today}

\begin{abstract}
	Universal quantum computers are potentially an ideal setting for simulating many-body quantum dynamics that is out of reach for classical digital computers. We use state-of-the-art IBM quantum computers to study paradigmatic examples of condensed matter physics -- we simulate the effects of disorder and interactions on quantum particle transport, as well as correlation and entanglement spreading. Our benchmark results show that the quality of the current machines is below what is necessary for quantitatively accurate continuous time dynamics of observables and reachable system sizes are small comparable to exact diagonalization. Despite this, we are successfully able to demonstrate clear qualitative behaviour associated with localization physics and many-body interaction effects. 
\end{abstract}

\maketitle


\section*{Introduction}

Quantum computers are general purpose devices that leverage quantum mechanical behaviour to outperform their classical counterparts by reducing the computational time and/or the required physical resources~\cite{Nielsen2010}. Excitement about quantum computation was initially fuelled by the prime-factorization algorithm developed by Shor~\cite{Shor1995}, which is most popularly associated with the ability to attack currently used cyber security protocols. More importantly, it provided a paradigmatic example of dramatic exponential improvement in computational speed when compared with classical algorithms. It has since been realised that the potential power of quantum computers could have far reaching applications, from quantum chemistry and the associated benefits for medicine and drug discovery~\cite{Kandala2017}, to quantum machine learning and artificial intelligence~\cite{Biamonte2017}. Even before reaching these lofty goals, there may also be practical uses for noisy intermediate-scale quantum (NISQ) devices~\cite{Preskill2018}.

These quantum devices can be implemented in a large number of ways, for example, using ultracold trapped ions~\cite{Cirac1995,Martinez2016,Nam2019,Wright2019,Bruzewicz2019,Figgatt2018}, cavity quantum electrodynamics (QED)~\cite{Houck2012,Barends2014,Blais2007,Zhu2017}, photonic circuits~\cite{Aspuru2012,Sun2018,Tambasco2018}, silicon quantum dots~\cite{Kane1998,West2019,Yang2019}, and theoretically even by braiding, as yet unobserved, exotic collective excitations called non-abelian anyons~\cite{Nayak2008,Sarma2006,Lahtinen2017}. One of the most promising approaches is using superconducting circuits~\cite{Nakamura1999,Wendin2017,Krantz2019}, where recent advances have resulted in devices consisting of up to 72 qubits, pushing us ever closer to realising so-called quantum supremacy~\cite{Boixo2018}. The apparent proximity of current devices to this milestone makes it timely to review the current capabilities and limitations of quantum computers.

Richard Feynman's original idea was to simulate quantum many-body dynamics -- a notoriously hard problem for a classical computer -- by using another quantum system~\cite{Feynman1982}. Over the last couple of decades this approach of using a purpose built \emph{quantum simulator} has been extremely successful in accessing physics beyond the reach of numerics on a classical computer. Most notable are cold atom experiments with optical lattices~\cite{Bloch2008,Bakr2009,Schreiber2015,Choi2016,Bordia2017,Mitra2017,Cooper2018} where the natural evolution of the atoms corresponds to a high accuracy to that of a local Hamiltonian of choice. This has, for example, allowed us to study Hubbard model physics~\cite{Bakr2009,Mitra2017} and many-body localized systems~\cite{Choi2016,Schreiber2015,Bordia2017} in two dimensions. More recently, there have also been advances in trapped ion quantum simulators, which have the benefit of being able to implement long range interactions~\cite{Cirac1995,Martinez2016,Garttner2017}, and have been used to study the Schwinger-mechanism of pair production/annihilation in 1D lattice QED~\cite{Martinez2016} and Floquet time crystals~\cite{Zhang2017}.

Universal quantum computers are also increasingly looking like a feasible setting for simulating quantum dynamics. One of the biggest advantages of using a quantum computer for this purpose is the flexibility it offers. A single quantum device could in principle perform simulations that currently require several different experiments, using disparate methods. Furthermore, it should be possible to access new physics not currently accessible, most notably, the simulation of lattice gauge theories with dynamical gauge fields. These are ubiquitous in the theoretical study of strongly-correlated quantum matter but require multi-body couplings, which have so far proven difficult to achieve in experiment~\cite{Wiese2013,Zohar2016}.

A particularly exciting opportunity has been provided by IBM in the form of an online quantum computing network called IBM Q. This consists of a set of small quantum computers of 5 and 16 qubits that are availably freely to the public, two 20 qubit machines accessible by IBM Q partners~\cite{IBMQ}, and the qiskit python API~\cite{qiskit} for programming the devices.
The publicly available resources have already resulted in a spread of results, such as calculating the ground state of simple molecules~\cite{OMalley2016,Kandala2017}, creating and measuring highly entangled many qubit states~\cite{Wang2018,Choo2018}, implementing quantum algorithms~\cite{Bravo-Prieto2019,Doronin2019,Amico2019}, and simulating non-equilibrium dynamics in the transverse-field Ising~\cite{Zhukov2018,Cervera-Lierta2018}, Heisenberg~\cite{Lamm2018} and Schwinger~\cite{Klco2018} models, as just a few examples. Given the infancy of quantum computing efforts, all these results are understandably small scale and of limited accuracy. 

IBM is not alone in their efforts to make quantum computing more mainstream, with Microsoft introducing the $Q$-sharp programming language~\cite{q-sharp}, Google developing the Circq python library~\cite{circq}, and Rigetti providing their own Quantum Cloud Service and Forest SDK built on Python~\cite{pyquil}. Rigetti and IonQ also provide selective public access to hardware -- based on superconducting qubits and trapped ions, respectively. All of these resources are allowing a lot of hands on experience with quantum computers from researchers and the general public all around the world. Furthermore, it highlights that there are currently several parallel efforts of research and development from industry focussed on quantum computing, quantum programming and cloud-based services that are flexible and prepared for future hardware. 

Given the current stage of development, and the immense expectation from the public and physics communities, it is timely to critically assess and benchmark the state-of-the-art. In this article we consider the far-from-equilibrium dynamics of global quantum quenches simulated on an IBM 20 qubit quantum computer. The models that we consider are of central importance to condensed matter physics and display a wide range of phenomenologies. By measuring a range of physical correlators, we can assess the capabilities and limitations of current quantum computers for simulating quantum dynamics.

\section*{Results}


\hfill\\\textbf{Setup}\\
In this article we study global quantum quenches~\cite{Essler2016,Vasseur2016}, that is we calculate local observables and correlators of the form
\begin{equation}\label{eq: observables}
\la \psi (t) | \hat{O}_j | \psi(t) \ra, \qquad \la \psi (t) | \hat{O}_j \hat{O}_k | \psi(t) \ra,
\end{equation}
where $\hat{O}_j$ are local operators and the time-dependent states are
\begin{equation}
|\psi(t) \ra = e^{-i \hat{H} t} |\psi(0)\ra,
\end{equation}
and where $|\psi(0)\ra$ is the initial state, which differs globally from an eigenstate of $\hat{H}$. In other words, we can consider $|\psi(0)\ra$ to be the ground state of a (time-independent) preparation Hamiltonian, $\hat{H}_0 = \hat{H} - h\hat{V}$, where $\hat{V}$ is a global perturbation, and the perturbation strength $h$ is instantaneously quenched from zero at time $t=0$.

We consider one dimensional spin-1/2 chains, consisting of $N$ spins, initially prepared in either a domain wall configuration, $|\cdots \!\downarrow \downarrow \downarrow \uparrow \uparrow \uparrow \!\cdots \ra$, or N{\'e}el state, $| \cdots \!\uparrow \downarrow \uparrow \downarrow \uparrow \!\cdots \ra$. Quenches from these initial states are particularly easy to prepare due to the local tensor-product structure of the initial state and have been studied in cold atom experiments~\cite{Schreiber2015}. The time evolution after the quantum quench will be governed by a Hamiltonian of the form
\begin{equation}\label{eq: H}
\hat{H} = - J\sum_{j=1}^{N-1} \left(\hat{\sigma}^x_j\hat{\sigma}^x_{j+1} + \hat{\sigma}^y_j \hat{\sigma}^y_{j+1}\right) + U \sum_{j=1}^{N-1} \hat{\sigma}^z_j \hat{\sigma}^z_{j+1} + \sum_{j=1}^N h_j \hat{\sigma}^z_j,
\end{equation}
with $J>0$, and $\hat{\sigma}^\alpha_j$ are the Pauli matrices with eigenvalues $\pm 1$. This model can also be written in terms of hard-core boson, or spinless fermion Hamiltonian via the Jordan-Wigner transformation. 

We will consider four distinct cases for the Hamiltonian parameters:
\begin{itemize}[noitemsep,topsep=3pt,parsep=3pt,partopsep=3pt,leftmargin=17pt]
	\item[\bf(i)] \label{case i} \textbf{The XX chain}, defined by $U=0$ and $h_j = 0$. This is a uniform, non-interacting model.
	\item[\bf(ii)] \label{case ii} \textbf{Disordered XX chain} for $U=0$ and $h_j$ uniformly randomly sampled from the interval $[-h,h]$. This is the prototypical lattice model for Anderson localization~\cite{Anderson1958}.
	\item[\bf(iii)] \label{case iii} \textbf{XXZ spin chain}, defined by $U\neq 0$ and $h_j = 0$. This is a particular case of the Heisenberg model for quantum magnetism.
	\item[\bf(iv)] \label{case iv} \textbf{XXZ chain with linear potential}, defined by $U > 0$ and $h_j = h j$, i.e., a linearly increasing potential. This model was recently studied in the context of many-body localization without disorder~\cite{Schulz2018,VanNieuwenburg}, where Wannier-Stark localization~\cite{Wannier1962} is induced by the linear potential.
\end{itemize}
This family of Hamiltonians covers a large range of physics in condensed matter, from the integrable limit~\cite{Essler2016}, to many-body quantum magnetism~\cite{Vasiliev2018}, to that of many-body localization~\cite{Nandkishore2015,Abanin2017}. These Hamiltonians are perfect testbeds for digital quantum simulation since they can be directly simulated on a quantum computer -- the spins-1/2 of the former correspond directly to the qubits of the latter, and the local connectivity of the qubits is suited for the local form of the Hamiltonian.

\begin{figure}[tb]
	\centering
	\subfigimg[width=.49\textwidth]{\textbf{(a)}}{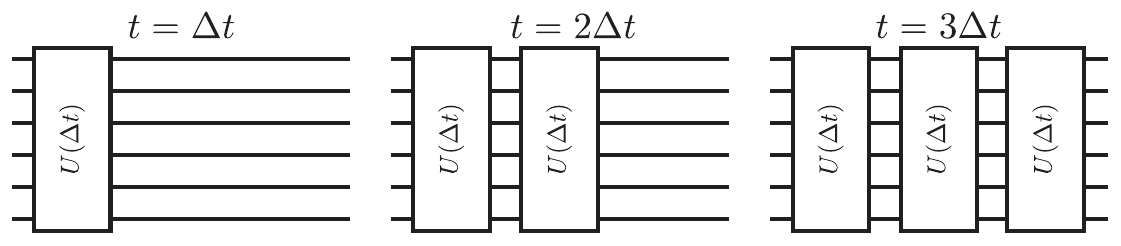}\\
	\subfigimg[height=.18\textwidth]{\hspace*{-4pt}\raisebox{-.6\baselineskip}{\textbf{(b)}}}{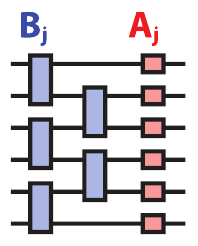}\quad
	\subfigimg[height=.18\textwidth]{\hspace*{-4pt}\raisebox{-.6\baselineskip}{\textbf{(c)}}}{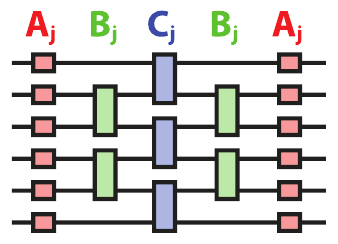}
	\caption{Schematic of the implementation of Trotterized evolution. (a) The procedure of adding more Trotter steps to reach longer times. (b) and (c) are the basic and symmetric Trotter decomposition, respectively, for the Hamiltonian Eq.~\eqref{eq: H}. In (b) the operators are $\hat{A}_j = e^{-i h_j \hat{\sigma}^z_j \Delta t}, \qquad \hat{B}_j = e^{-i(U\hat{\sigma}^z_j\hat{\sigma}^z_{j+1}- J(\hat{\sigma}^x_j\hat{\sigma}^x_{j+1} +\hat{\sigma}^y_j\hat{\sigma}^y_{j+1}))\Delta t}$ and in (c) they are $\hat{A}_j = e^{-i h_j \hat{\sigma}^z_j \frac{\Delta t}{2}}, \quad \hat{B}_j = e^{-i(U\hat{\sigma}^z_j\hat{\sigma}^z_{j+1}-J(\hat{\sigma}^x_j\hat{\sigma}^x_{j+1} +\hat{\sigma}^y_j\hat{\sigma}^y_{j+1}))\frac{\Delta t}{2}}, \hat{C}_j = e^{-i(U\hat{\sigma}^z_j\hat{\sigma}^z_{j+1}-J(\hat{\sigma}^x_j\hat{\sigma}^x_{j+1} -\hat{\sigma}^y_j\hat{\sigma}^y_{j+1}))\Delta t}$.}\label{fig: trotter}
\end{figure}

In our simulations we achieve the time evolution using a Trotter decomposition of the unitary time evolution operator $\hat{U}(t) = e^{-i\hat{H}t}$. Our simulation proceeds by the following steps. First, we create the initial state, which is done by applying Pauli-X operations to the default initial state of the IBM machine, $|\cdots \uparrow\uparrow\uparrow \cdots \ra$. Second, we discretize time and the evolution becomes a product of discrete evolution operations $\hat{U}(t) = \hat{U}(\Delta t) \cdots \hat{U}(\Delta t)$. Note that in our plots we include additional data points by varying $\delta t$ in the final discrete evolution step. Third, we trotterize the discrete operators $U(\Delta t)$ into a product of one- and two-qubit unitaries. Finally we decompose these unitaries into the CNOT and single qubit rotation gates that can be directly applied on the IBM devices. Please see Fig.~\ref{fig: trotter} for a schematic of this procedure, and see Methods for more details.


Once we have constructed the state $|\psi(t)\ra$, we then perform measurements, which allows us to construct the quantities of interest in Eq.~\eqref{eq: observables}. We will only consider correlators of $\hat{\sigma}^z_j$ operators, where we only need to make measurements in a single ($z$-basis) for the spins. In the following we will use 8192 measurements per data point, which means that the statistical error for these local correlators is $\sim 0.01$, which is too small to be included in our figures.

We consider three different types of dynamical quantities:
\begin{itemize}[noitemsep,topsep=3pt,parsep=3pt,partopsep=3pt,leftmargin=10pt]
	\item The local magnetization 
	\begin{equation}
	M_j(t) = \la \psi (t) | \hat{\sigma}^z_j | \psi(t) \ra.
	\end{equation}
 	In the case of a domain wall initial state, we will also compute
	\begin{equation}\label{eq: N half}
	N_\text{half}(t) = \sum_{j=1}^{N/2} \la \psi (t) |\, \frac{\hat{\sigma}^z + 1}{2} \,| \psi(t) \ra,
	\end{equation}
	which is equal to zero at $t=0$ and grows as the domain wall spreads.
	\item The connected equal-time correlator
	\begin{equation}\label{eq: connected correlator}
	\;\; C_{jk}(t) = \la \psi (t) | \hat{\sigma}^z_j\hat{\sigma}^z_k | \psi(t) \ra - \la \psi (t) | \hat{\sigma}^z_j| \psi(t) \ra\la \psi (t) | \hat{\sigma}^z_k | \psi(t) \ra.
	\end{equation}
 	Note, the connected form of this correlator measures the quantum correlators between distant spins but is not sensitive to the classical correlations in our initial states.
	\item The quantum Fisher information (QFI). We will consider a particular case of the QFI for a pure state, namely,
	\begin{equation}\label{eq: QFI}
	\;\; F_Q(t) = \sum_{jk} s_j s_k \la \psi (t) | \hat{\sigma}^z_j\hat{\sigma}^z_k | \psi(t) \ra - \left(\sum_j s_j \la \psi (t) | \hat{\sigma}^z_j | \psi(t) \ra \right)^2,
	\end{equation}
	where $s_j = +1$ for left half of the sites $j$ and $s_j = -1$ for the right half at $t=0$. More generally, the QFI (for a pure state) is defined as the variance, $4\left(\la \hat{O}^2\ra - \la\hat{O}\ra^2\right)$, where $\hat{O}$ is a sum of local operators, which each have a spectrum of unit width. In our case we have $\hat{O} = \frac{1}{2}\sum_j s_j \hat{\sigma}^z_j$. The QFI is an entanglement witness~\cite{Hyllus2012,Toth2012,Hauke2016}, and for our chosen definition in Eq.~\eqref{eq: QFI} is also closely related to the von Neumann entanglement entropy~\cite{Klich2009,Song2011,Song2012}.
\end{itemize}

\hfill\\\textbf{The IBM Quantum Computers}\\
The quantum computer that we use is the latest 20-qubit IBM device, codenamed \texttt{ibmq\_poughkeepsie}. It consists of a two-dimensional array of qubits that have local connectivity. We can perform arbitrary single qubit rotations and controlled-NOT (CNOT) gates between connected qubits, see Methods. For the data presented in this paper the average readout errors, CNOT errors, and T2 (dephasing) times were approximately $4\%, 2\%$ and $90\mu$s, respectively. An important point to note is that the IBM machines are recalibrated on an approximately daily basis, which means the data can vary across days. Crucially, we find that the our results are qualitatively reproducible, and we compare data obtained across three consecutive days in the Methods. 

To benchmark the accuracy of the simulation, we compare the data with a numerical implementation of the Trotter evolution, as well as continuous-time exact diagonalization (ED). The errors in our results are strongly influenced by the number of CNOT gates in the corresponding quantum circuit. One of the reasons for this is that the fidelities of these two qubit gates are an order of magnitude worse than the single qubit gates. The CNOTs also take a longer real-world time to implement than the single qubit gates. The increased implementation time of the circuits increases the potential for errors due to energy relaxation and dephasing, parametrized by the T1 and T2 times, as well as other environmental effects and cross-talk. On the IBM device, we use $N=6,8,10$ of the qubits as our system with 4 or 5 symmetric Trotter steps. These qubits are chosen as the connected subset with the lowest average CNOT errors such that the single qubit measurement errors and T2 decoherence times do not exceed a certain threshold. Please see Methods for more details about the quantum device and details of the algorithm used to select the qubits. All data presented below is available in Ref.~\cite{IBMdata}.

\begin{figure*}[t]
	\centering
	\subfigimg[height=.235\textwidth]{\hspace*{10pt}\textbf{(a)}}{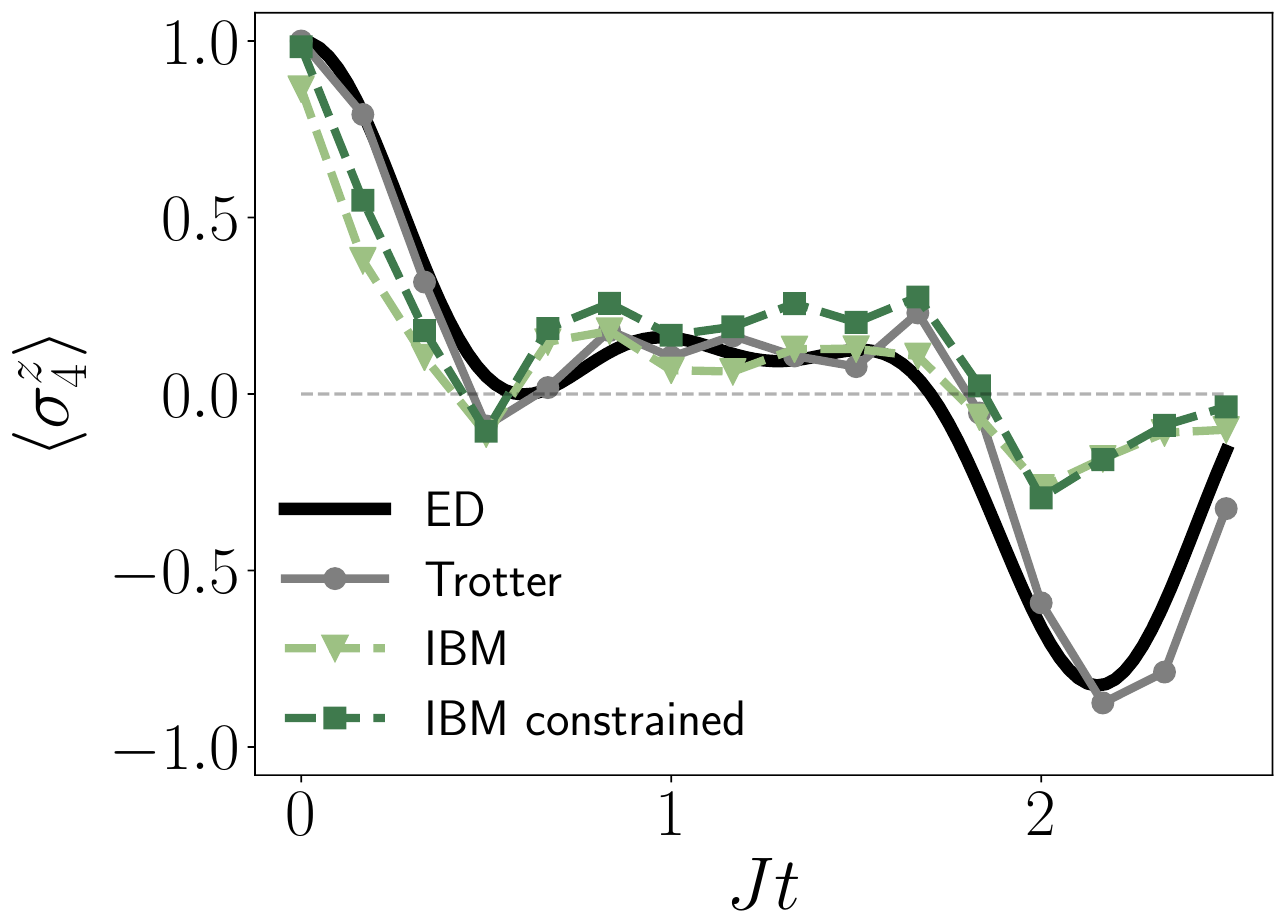}
	\subfigimg[height=.235\textwidth]{\hspace*{10pt}\textbf{(b)}}{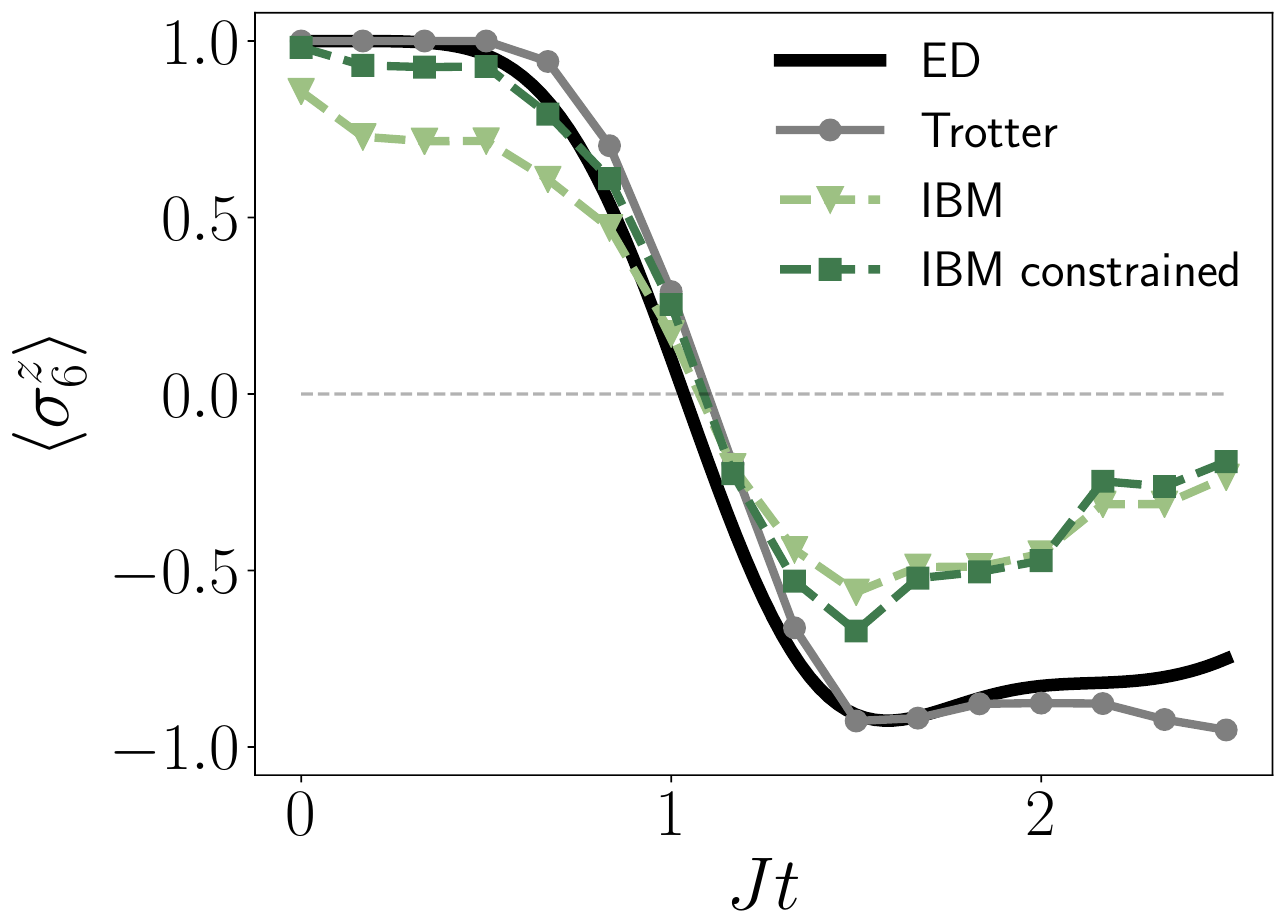}
	\subfigimg[height=.235\textwidth]{\hspace*{10pt}\textbf{(c)}}{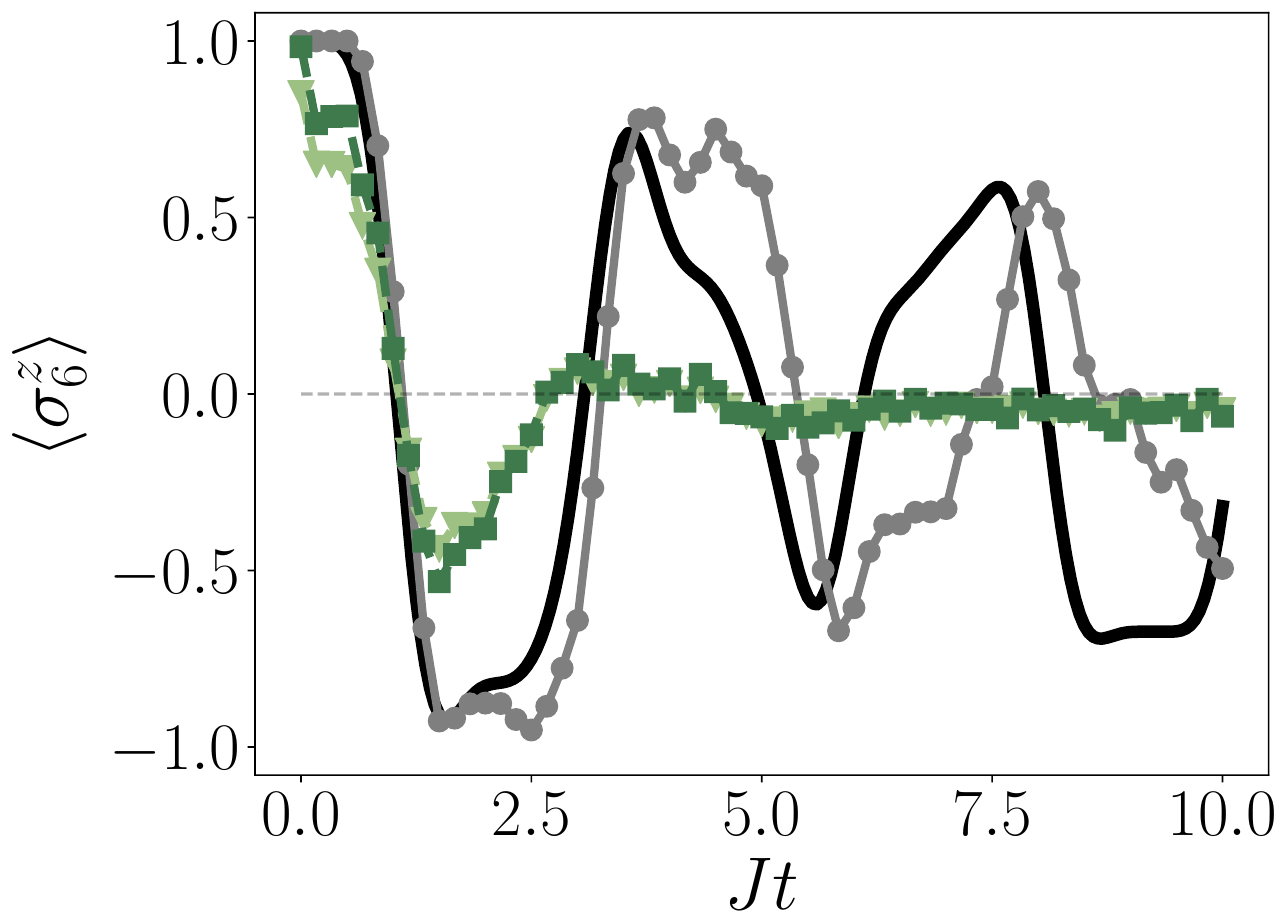}\\
	\qquad\subfigimg[height=.27\textwidth]{\textbf{\quad(d)}}{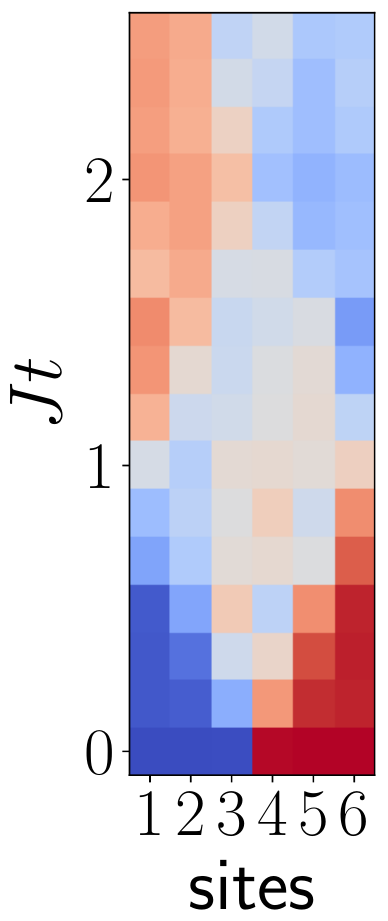}
	\;\subfigimg[height=.27\textwidth]{\textbf{(e)}}{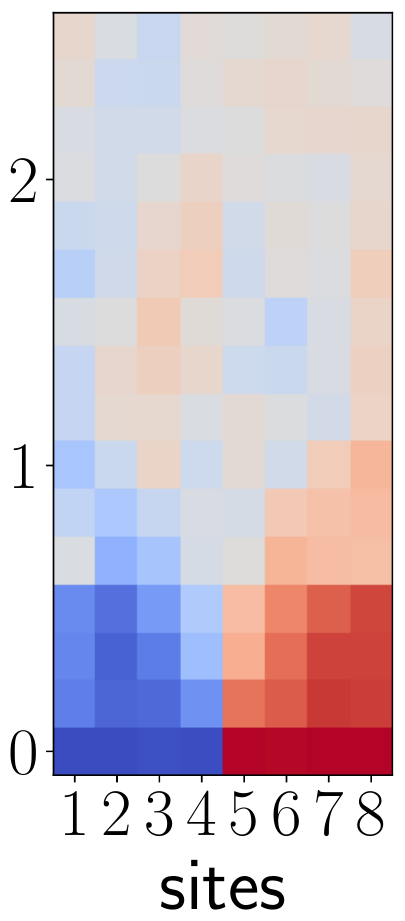}
	\;\subfigimg[height=.27\textwidth]{\textbf{(f)}}{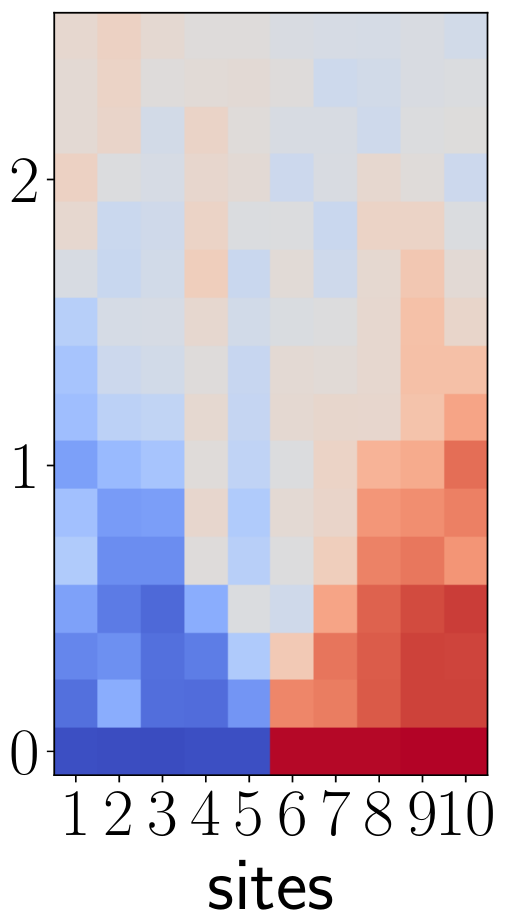}
	\;\subfigimg[height=.274\textwidth]{\textbf{(g)}}{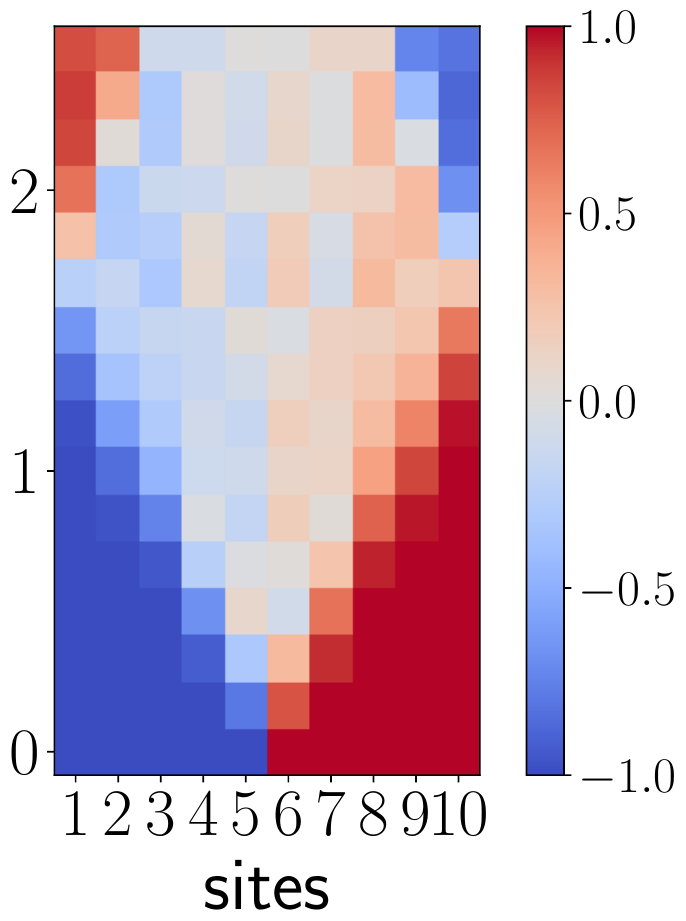}
	\caption{Results for a global quench from a domain wall initial state. (a-b) The local magnetization of the fourth and sixth spins of the chain with $N=6$, and (c) is for the sixth spin for longer times. Plotted are the result of exact diagonalization, numerical Trotter evolution, and the experimental data, both in raw form (IBM) and when the conserved quantities are imposed (IBM constrained), see the main text. (d-f) The time and site resolved IBM results for the local magnetisation for $N=6,8,10$ sites, respectively. (g) The corresponding numerical symmetric Trotter evolution for $N=10$. Data was obtained on 12 March 2019 for all figures except (c), which was obtained on 3 April 2019.
	}\label{fig: tests}
\end{figure*}

\begin{figure}[t]
	\centering
	\hspace*{-200pt}\textbf{(a)}\\
	{\rowcolors{2}{white}{gray!15}
		\setlength{\arrayrulewidth}{1pt}
		\setlength{\tabcolsep}{0pt}
		\renewcommand{\arraystretch}{1.2}
		\begin{tabular}{| l c | c | c | c | c |}
			\hline
			\;Date \;&\; \;&\; 8 May \;&\; 9 May \;&\; 10 May \;&\; 11 May \;\\
			\hline
			\;$M_\text{GHZ}$ \;&\; \;&\; 2.9597 \;&\; 2.4734 \;&\; 2.3977 \;&\; 2.9705 \;\\
			\hline
	\end{tabular}}\\\vspace*{7pt}
	\subfigimg[width=.24\textwidth]{\hspace*{10pt}\textbf{(b)}}{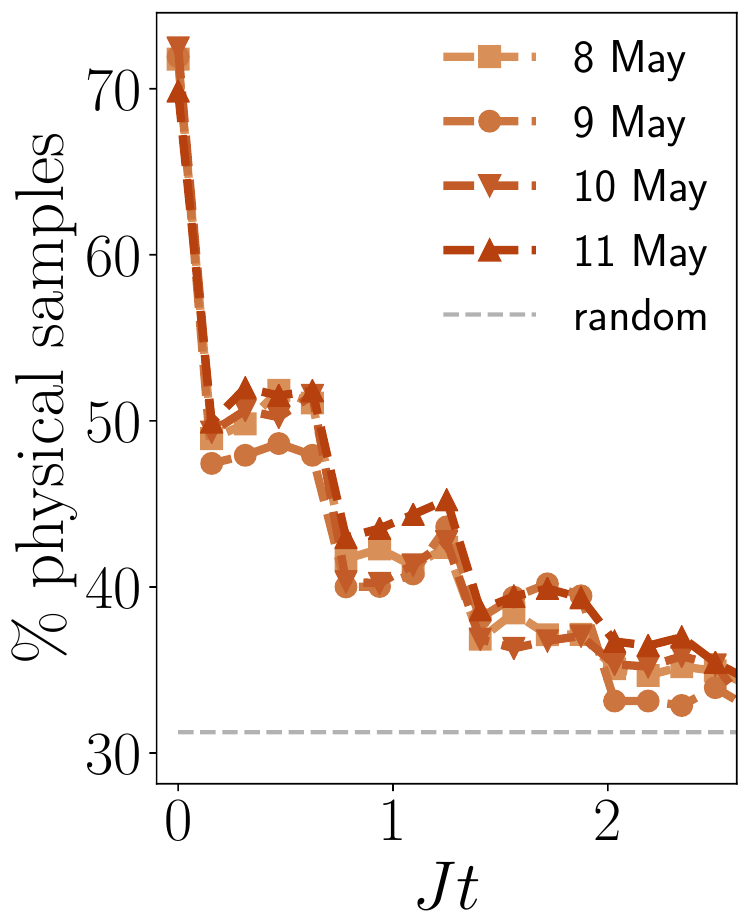}
	\!\subfigimg[width=.24\textwidth]{\hspace*{10pt}\textbf{(c)}}{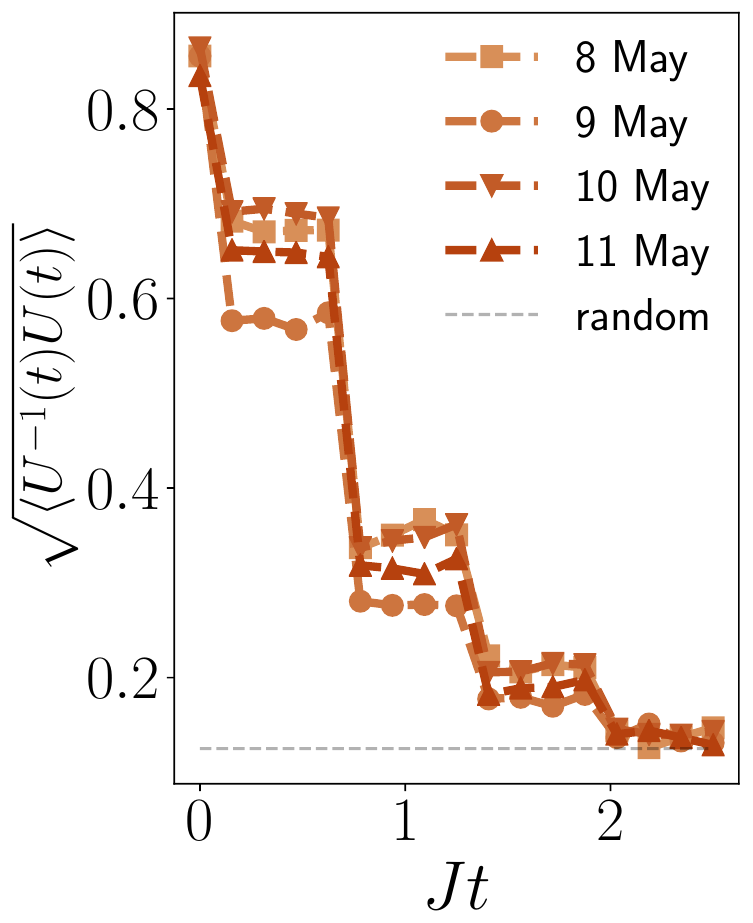}
	\caption{(a) Values of $M_\text{GHZ}$ defined in Eq.~\eqref{eq: GHZ M} computed on the IBM device for a GHZ state on three qubits. Values of $M_\text{GHZ}>2$ cannot be explained by a classical theory of local realism, see main text. (b) The percentage of measured states that are in the physical Hilbert space as a function of time after a quench from a domain wall. (c) Data for the computation of the quantity $\sqrt{\hat{U}^{-1}(t) \hat{U}(t)}$, as performed on the IBM device, after a quench from a N{\'e}el state. Data was obtained for $N=6$ across four consecutive days in 2019. Dashed lines indicate the average value for a randomly selected state.}\label{fig: accuracy tests}
\end{figure}

\hfill\\\textbf{Local Magnetization}\\
First, we consider results for the uniform XX spin chain Hamiltonian with $U=0$ and $h_j=0$ (\hyperref[case i]{case (i)}), quenched from a domain wall configuration, shown in Fig.~\ref{fig: tests}. Figure~\ref{fig: tests}(a-c) show a comparison of the magnetization of the fourth (middle) and sixth (end) spins of the chain as computed by exact diagonalization with continuous time evolution, a numerical implementation of the Trotter decomposition and the corresponding data from the IBM machine. The data from the machine is further split into the raw data (orange triangles) and constrained data (red squares). The constrained data only considers those measurement outcomes that have the same total magnetization as the initial state -- which is a conserved quantity -- that is, we restrict to the physical Hilbert space of the Hamiltonian we are simulating. We discuss this rudimentary error mitigation method further in the context of quantifying the accuracy of the quantum computer (see Fig.~\ref{fig: accuracy tests}(b)), and it will be used in all subsequent figures. See Methods for more details.

The data in Fig.~\ref{fig: tests}(a-b) show that while all curves show reasonable agreement at short times -- for instance, we have a delayed decay of the magnetization in Fig.~\ref{fig: tests}(b) -- the accuracy of the IBM data becomes bad very quickly. For times $Jt > 1$ the magnetization as measured on the machine approaches zero, which is the expected average value if the system thermalizes or if we were to randomly sample states. The agreement between the numerical results obtained from exact diagonalization (ED) and trotterization shows that the inaccuracy of the results is due to the machine and not our approximation of the evolution. This rapid decay in the accuracy with number of Trotter steps was also observed in Ref.~\cite{Zhukov2018} for the transverse field Ising model on the 5 and 16 qubit IBM machines~\cite{Leeb2018}.

In Fig.~\ref{fig: tests}(c) we consider the evolution over a longer time window, up to $Jt=10$, rather than $Jt=2.5$. We see that the data very quickly approaches $0$ and remains there, indicating that the system has equilibrated. From this figure we can see that due to the quality of the machine we want to take $\Delta t$ in our Trotter steps as large as possible so that we can reach longer times without thermalizing. However, for the Trotter evolution to be accurate we want $\Delta t$ as small as possible. The $\Delta t$ that we use are chosen (without much optimization) to maintain a reasonable accuracy in both the IBM data and the numerical trotterization.

Next, we consider the site dependence of the magnetization shown in Figs.~\ref{fig: tests}(d-f), for $N=6,8,10$ sites and compared with the numerically computed Trotter evolution shown in Fig.~\ref{fig: tests}(g). In these figures we see a clear qualitative agreement between the experimental and numerical results at short times, particularly by the presence of a linear light-cone causality structure for the spreading of the domain wall. This qualitative agreement, however, also worsens at longer times, and as we increase the system size. In particular, in Fig.~\ref{fig: tests}(d) we can see a marked decrease in accuracy every three time steps. The origin can be explained as follows. Each block of three time steps is computed for the same number of time steps with $\Delta t$ varying in the final Trotter steps (as explained earlier). It is when we add an additional time step for the next block -- and thus increase the number of gates in the quantum circuit -- that we see a drop in the accuracy. This behaviour is also seen in Fig.~\ref{fig: accuracy tests}(b) where we show the number of measurements that are in the physical Hilbert space. There is a clear decrease in the percentage after the introduction of each new Trotter step.

\begin{figure*}[t]
	\centering
	\subfigimg[width=.33\textwidth]{\hspace*{10pt}\textbf{(a)}}{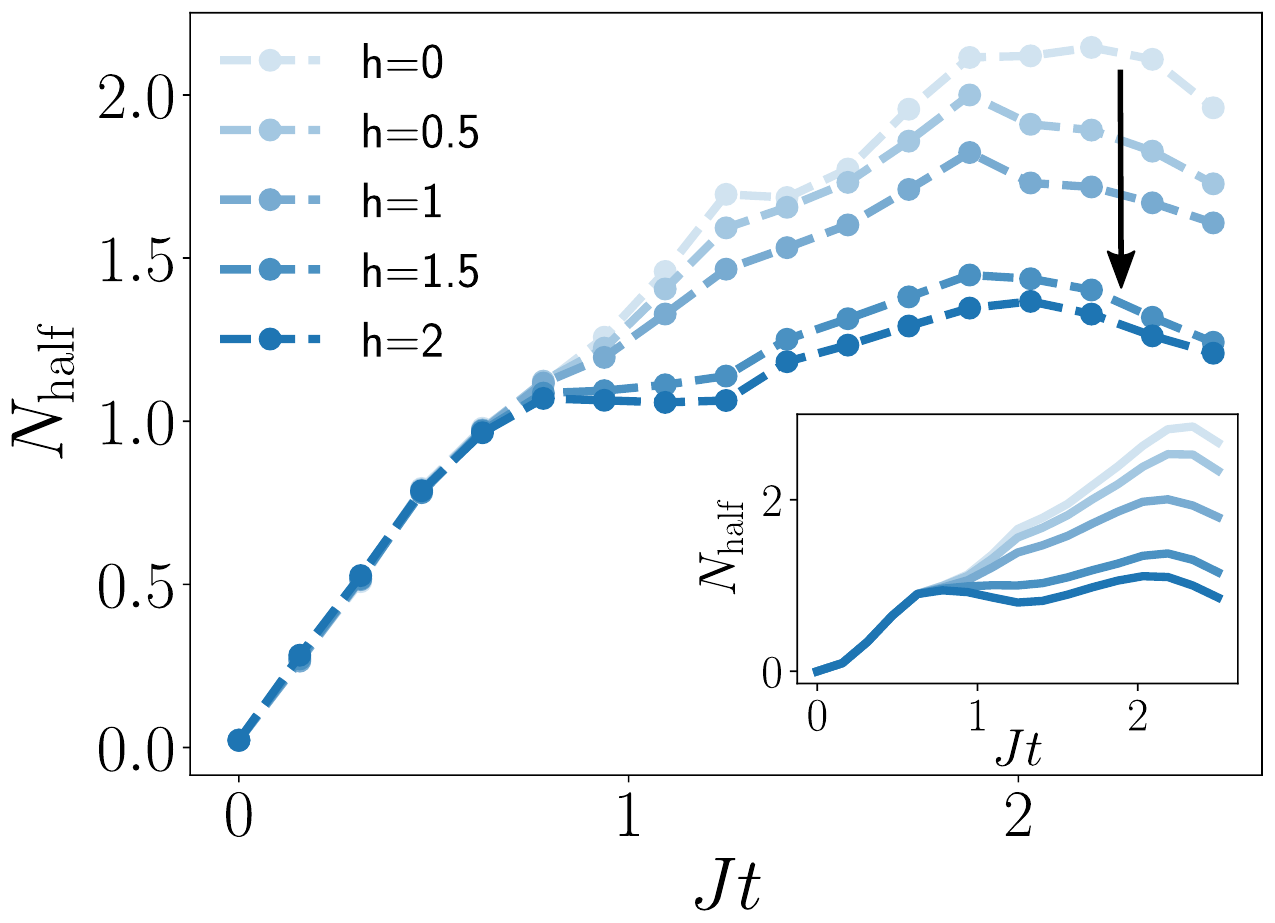}
	\subfigimg[width=.33\textwidth]{\hspace*{10pt}\textbf{(b)}}{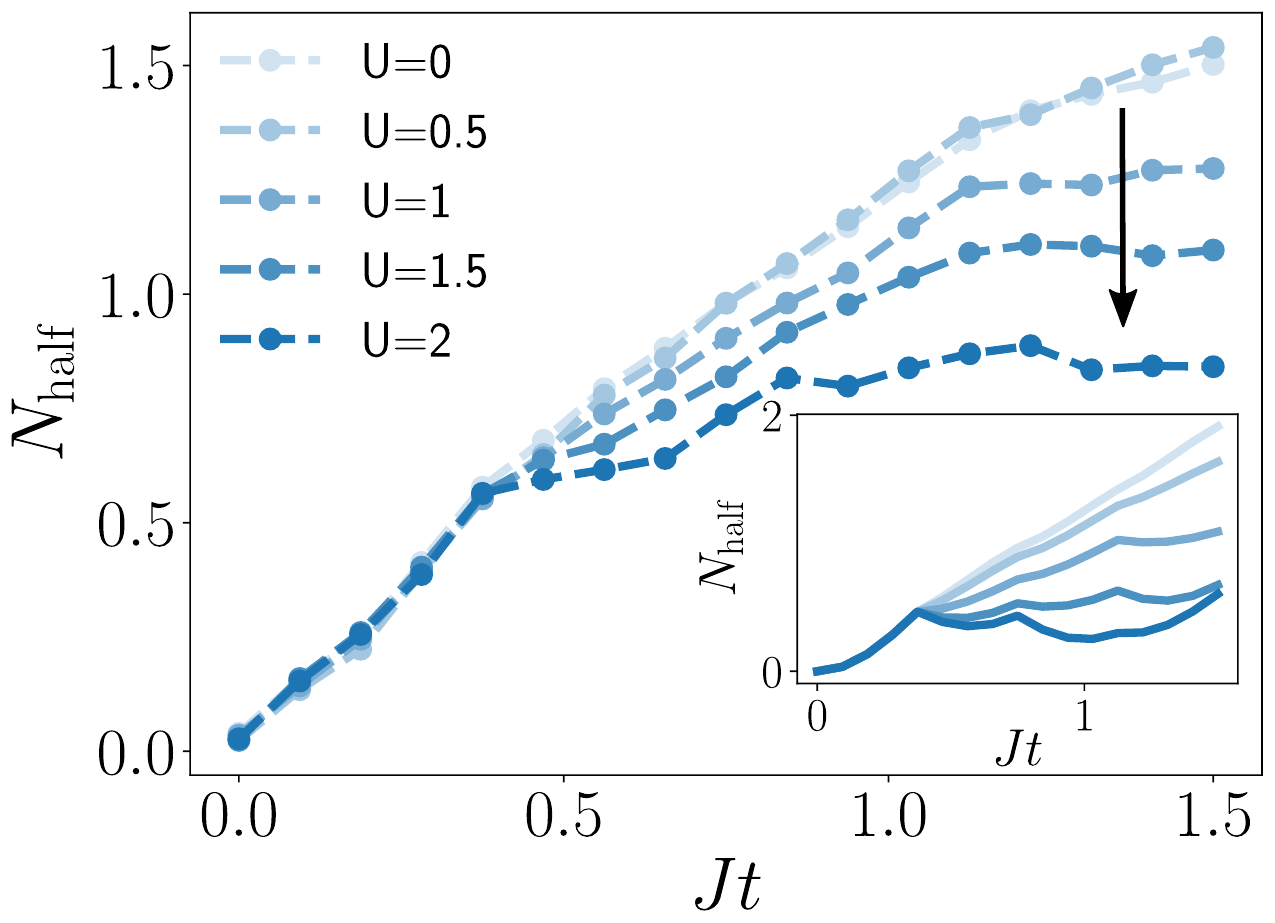}
	\subfigimg[width=.33\textwidth]{\hspace*{10pt}\textbf{(c)}}{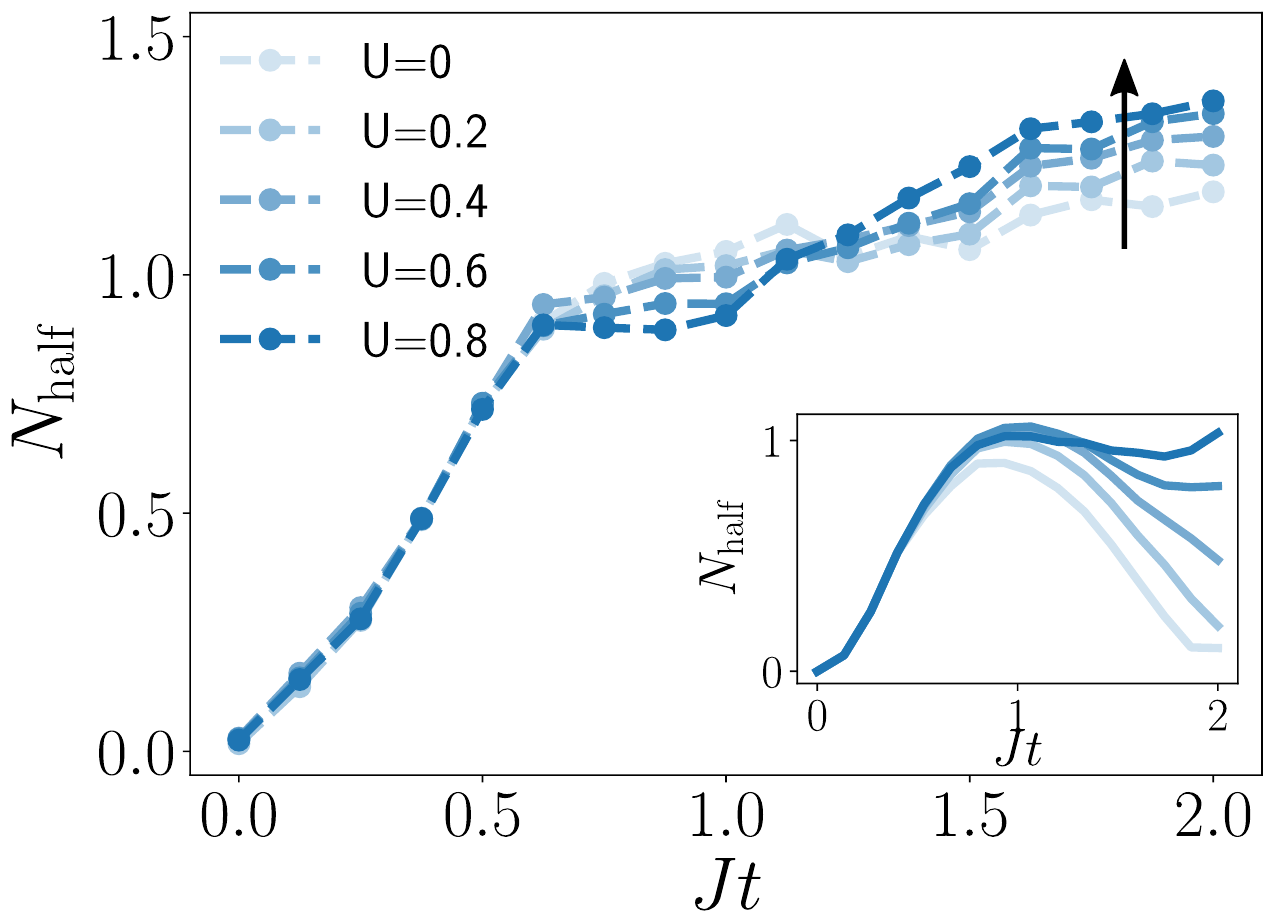}
	\caption{Results for short time behaviour of $N_\text{half}$ after a quench from a domain wall initial state. In all subfigures the inset shows the corresponding results of numerical Trotter evolution. (a) The disordered XX spin chain with disorder strength controlled by $h$ using a symmetric Trotter decomposition. (b) The XXZ spin chain with nearest neighbour interactions parametrized by $U$ with basic trotterization. (c) The XXZ spin chain with a linear potential with slope $h=1.5$, interaction strength $U$, and a symmetric Trotter decomposition. (a-b) were obtained on 12 March 2019 and (c) on the 10 May 2019.}\label{fig: disorder}
\end{figure*}


\hfill\\\textbf{Short-Time Many-Body Physics}\\
While the results in Fig.~\ref{fig: tests} may at first seem discouraging for \emph{quantitative} large-scale dynamical simulations, we will show in this section that we may still observe \emph{qualitative} behaviour associated with non-trivial quantum phenomena. Here, we consider $N_\text{half}(t)$, defined in Eq.~\eqref{eq: N half}, after quenching from the domain wall initial state for three different cases.

First, we consider the disordered XX chain (\hyperref[case ii]{case (ii)}), which is known to exhibit Anderson Localization in 1D for all values of the disorder strength, $h$~\cite{Kramer1993}. As a consequence of increasing the disorder strength, the extent of the spreading of the domain wall, and consequently the growth of $N_\text{half}$ is reduced (indicated by a black arrow). We are able to reproduce this behaviour qualitatively for short times on the IBM machine as shown in Fig.~\ref{fig: disorder}(a). The corresponding numerical Trotter results are shown in the inset, which shows that while the accuracy of the results is quite low, the qualitative behaviour is still captured. Once again, we see that the data is biased towards the scrambled value as the number of Trotter steps is increased, which in the present case is $1.5$.

\begin{figure}[b]
	\centering
	\subfigimg[height=.36\textwidth]{\textbf{\;\qquad(a)\, Exact}}{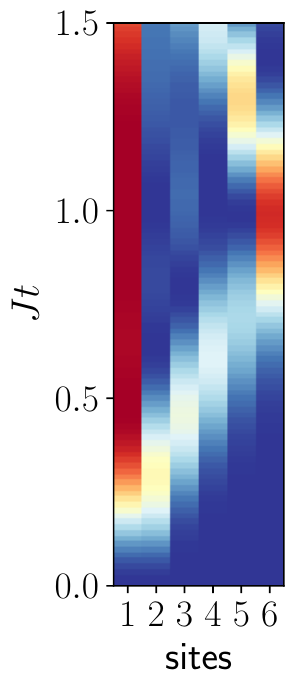}
	\!\!\!\subfigimg[height=.354\textwidth]{\raisebox{2pt}{\textbf{\quad\!(b)\, Trotter}}}{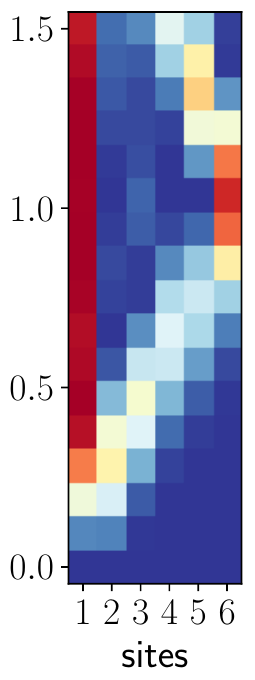}
	\!\!\!\subfigimg[height=.36\textwidth]{\textbf{\quad\!(c)\; IBM}}{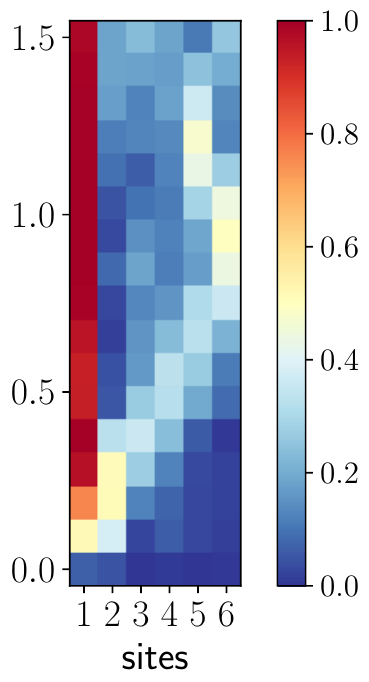}
	\caption{Results for the connected spin correlator defined in Eq.~\eqref{eq: connected correlator}. Data is computed using: (a) exact diagonalization, (b) numerical trotterization, (c) the IBM quantum device. Data is shown for $N=6, h_j=0$ and $U=0$, using a symmetric trotterization, and obtained on 12 March 2019. }\label{fig: correlators}
\end{figure}

Next, we consider the uniform XXZ chain (\hyperref[case iii]{case (iii)}) with $U>0$ and $h_j = 0$. In contrast to the previous cases, this Hamiltonian is interacting and thus describes true many-body physics. At short times, the spreading of the domain wall is also hindered by the energy cost of the additional interactions between neighbouring spins. Once again, we can see this behaviour qualitatively in the experimental results, shown in Fig.~\ref{fig: disorder}(b). Note that while the short time results are similar to the previous case, it is due to a different physical mechanism and is a many-body effect. The long-time behaviour would be starkly different from that of the Anderson localized case, which is, however, beyond the current capabilities of the IBM quantum computers.

In the third case we combine features of the previous two models and include both on-site potential energies and interactions. We will consider the XXZ spin chain with a linear potential (\hyperref[case iv]{case (iv)}), and both $U>0$ and $h>0$. If we compare with $U=0$, that is, a linear potential alone, then the eigenstates will be localized~\cite{Schulz2018,Wannier1962,VanNieuwenburg}, and thus the spreading of the domain wall will be limited. If we have $U>0$, then the two energy costs can compensate. For instance, consider flipping the middle two spins, then there will be an increase in energy due to the potential but a decrease in the interaction energy. Therefore, the presence of interactions makes it easier for the domain wall to spread resulting in an increase of $N_\text{half}$ as a function of $U$ (see black arrow). This simple argument of energetics is confirmed by the numerical results in the inset of Fig.~\ref{fig: disorder}(c), and is qualitatively reproduced in the experimental data shown in the main figure. Note, however, that this trend is less pronounced than in the previous two cases. This is in part due to the small scale of the changes in the exact results, as well as the bias towards the thermal value of 1.5 at long times.

In this data, the addition of disorder and interactions lead to similar qualitative behaviour on the time scales that we have considered. However, at longer times there is a clear difference between the two cases. In the former we have localization behaviour leading to the long-time persistence of the initial imbalance, whereas interactions generically lead to ergodic and thermalizing behaviour, resulting in the loss of this information in local observables. With the current devices we are unfortunately unable to distinguish these two different regimes.


\hfill\\\textbf{Spreading of Correlations -- Light Cone}\\
In the previous section, all the data correspond to some combination of local average magnetizations. Here, we look at the correlations between pairs of spatially separated spins, and how these correlations spread. Lieb and Robinson showed that for a local Hamiltonian, correlations can spread at most linearly and display a light-cone causality structure~\cite{Lieb1972}. For instance, for tight-binding spinless fermions, correlations spread with a speed proportional to their maximum group velocity, $v=4J$~\cite{Essler2016}.

In Fig.~\ref{fig: correlators} we compare ED, numerical trotterization and experimental results for the connected spin correlator defined in Eq.~\eqref{eq: connected correlator}, for $h_j = U = 0$. We measure correlations between the first and the $j^\text{th}$ spin after a quench from a charge density wave initial state. The ED results show a clear ballistic spreading of correlations, which is also qualitatively reproduced by the numerical trotterization and the IBM data. As with all previous results, the agreement between the numerical and IBM results is best at short times, but the IBM data is able to capture the point where the correlations reach the system size. Furthermore, for this free model, there are only significant correlations along the light-cone and not within it, which is also approximately captured by the experimental data. There does, however, seem to be a slightly faster light-cone velocity, which is most evident in the shift of the position of the peak on site 6. This indicates that there is an effective renormalization of the Hamiltonian parameters, particularly $J$, due to the errors in the machine.

\begin{figure}[t]]
	\centering
	{\hspace*{-10pt}\subfigimg[height=.29\textwidth]{\hspace*{10pt}\textbf{(a)}}{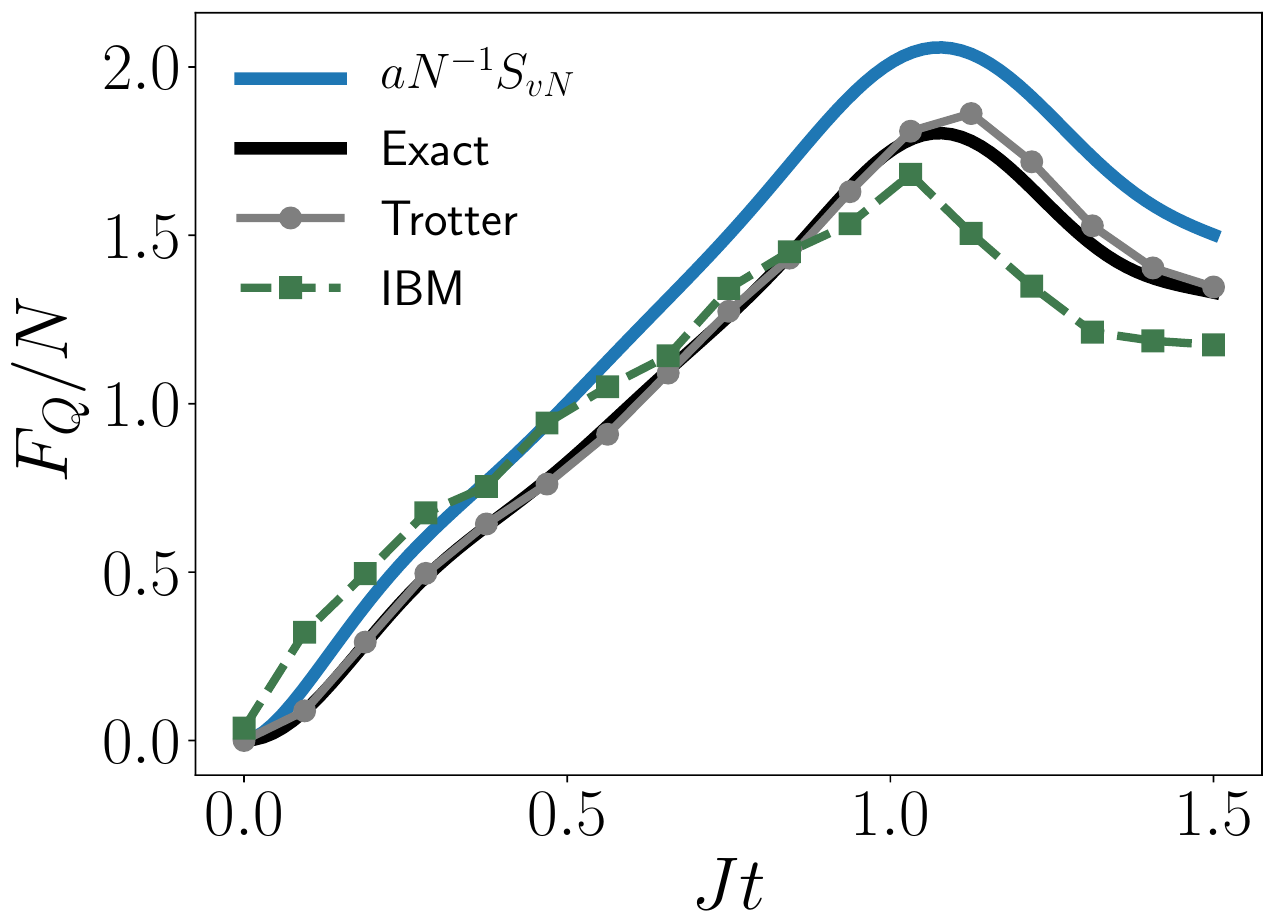}}
	{\hspace*{-10pt}\subfigimg[height=.29\textwidth]{\hspace*{10pt}\textbf{(b)}}{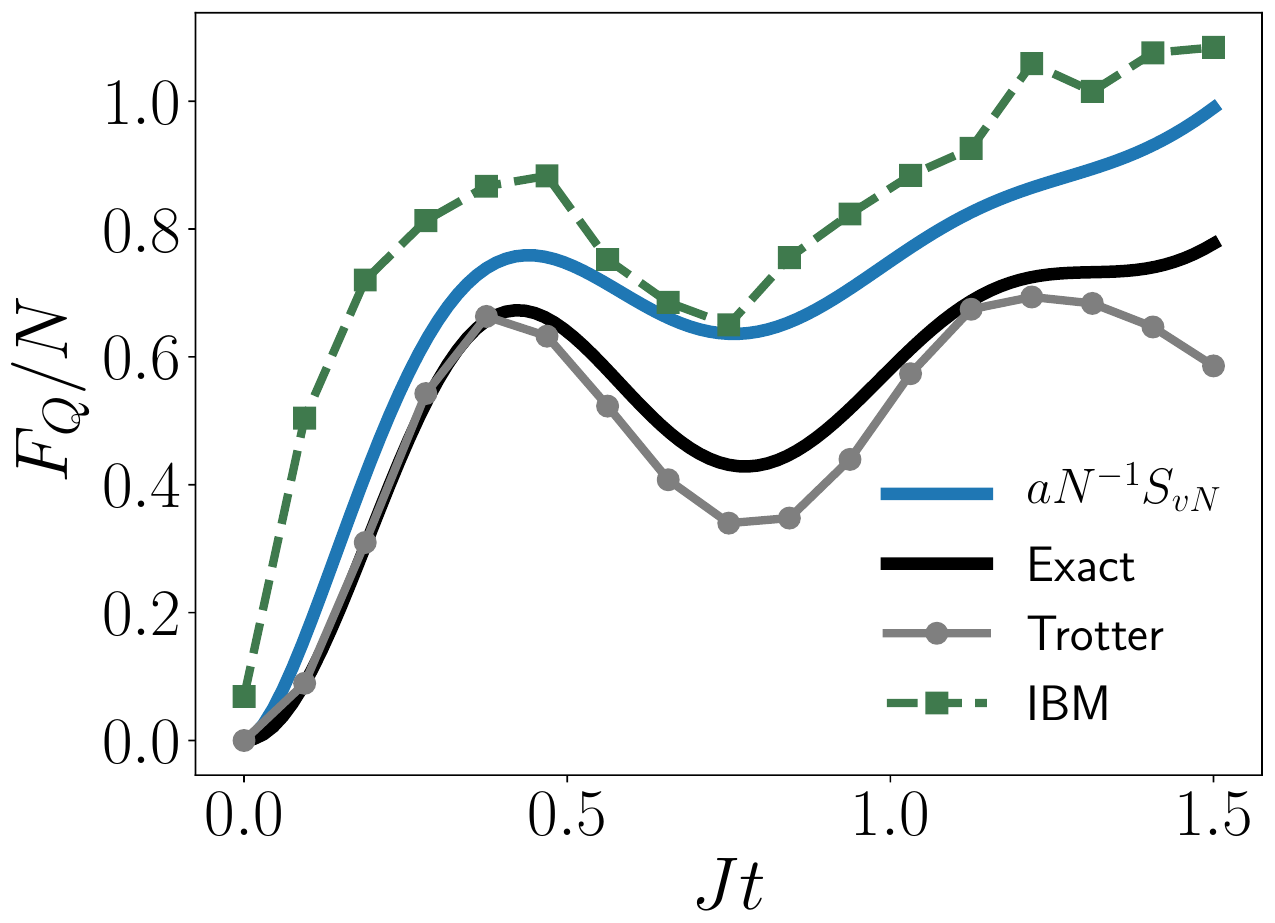}}
	\caption{The quantum Fisher information as computed by ED, numerical trotterization and using the IBM quantum device for $h_j=0$ and $U=0$ and a symmetric trotterization. Data is compared with the bipartite von Neumann entanglement entropy $S_\text{vN}$ with an equal left/right partition, scaled by $aN^{-1}$ with $a=32/5$, see main text. (a) Quench from the N{\'e}el initial state. (b) Quench from the domain wall initial state. Data was obtained on 12 March 2019.}\label{fig: QFI}
\end{figure}

\hfill\\\textbf{Quantum Fisher Information}\\
Finally, we consider the quantum Fisher information defined in Eq.~\eqref{eq: QFI} starting from both the N{\'e}el and domain wall initial states, shown in Figs.~\ref{fig: QFI}(a) and (b) respectively. In this section we will consider the model of (\hyperref[case i]{case (i)}), for which the quantum Fisher information has two important relations to entanglement, which we outline below. Note that the quantum Fisher information has a more general definition for mixed states, which reduces to our definition for pure states. We do not consider the more general definition since we are simulating unitary evolution from a pure quantum state and we are not able to differentiate between the unitary and non-unitary errors occurring in our circuits. 

For non-interacting models (as is the case for the Hamiltonian of (\hyperref[case i]{case (i)})), there is a direct relationship between the bipartite von Neumann entanglement entropy and the magnetization fluctuations~\cite{Klich2009,Song2011,Song2012}. The former is defined by $S_\text{vN} = - \text{Tr}[\rho^A \ln \rho^A]$, where $\rho^A$ is the reduced density matrix for half of the system.  
The variance of the half chain magnetization is proportional to our definition of the QFI and we have the approximate relation
\begin{equation}
S_\text{vN} \approx \frac{5}{32} F_Q,
\end{equation}
see Ref.~\cite{Song2012} for details of this cumulant expansion. In MBL systems the QFI -- using instead the staggered magnetization for the operator $\hat{O}$ in Eq.~\eqref{eq: QFI} -- also appears to mimic the bipartite entanglement entropy and grows logarithmically after a quantum quench~\cite{Smith2016a}.

The QFI is also a multi-partite entanglement witness~\cite{Pezze2009,Toth2012,Hyllus2012}, and in particular, if the state is separable then we have that $F_Q \leq N$, which is known as the shot noise limit in quantum metrology~\cite{Helstrom1976,Holevo1982,Giovannetti2004}. For a general entangled state, however, the QFI is bounded by $F_Q \leq N^2$, and a value of $F_Q/N \geq m$, indicates at least $m+1$-partite entanglement. Note that the QFI is sensitive to the choice of the operator $\hat{O}$, and for example, if we choose $\hat{O} \propto \sum_{j} \hat{\sigma}^z_j$, the total magnetization, then $F_Q = 0$, since our models conserve the total magnetization.

Let us now consider the IBM results for the QFI, starting with a quench from a N{\'e}el initial state shown in Fig.~\ref{fig: QFI}(a). We first note that the numerical results for the QFI (black) closely follow the bipartite von Neumann entanglement entropy (blue). The results from the IBM machine are also able to reproduce this behaviour quite accurately, characterized by the linear growth with a maximum just after $Jt=1$ due finite size effects. The fact that we measure $F_Q/N > 1$ also implies entanglement in the state on the IBM quantum computer. 

In Fig.~\ref{fig: QFI}(b) we consider a quench from a domain wall. In this case both the numerical and experimental simulations have the same qualitative behaviour but are further off in absolute value, as compared with Fig.~\ref{fig: QFI}(a). The QFI computed on the IBM machine once again is able to reproduce the behaviour of the von Neummann entanglement entropy.

The entanglement entropy is generally a difficult quantity to measure. It typically requires some form of state tomography, which consists of a set of measurements in a number of different bases that grows exponentially with system size, rendering it impractical for large systems. Furthermore, even with low error rates, the resulting density matrix may be unphysical~\cite{Smolin2012,Choo2018}. The quantum Fisher information may provide an alternative in certain circumstances because it can be significantly easier to compute -- for the definition that we consider we only need to measure in a single basis.


\hfill\\\textbf{Quantifying the Accuracy of the Quantum Computer}\\
While in the previous sections we showed that the current IBM device can qualitatively reproduce physical behaviour, it is also important to develop practical quantitative measures of their accuracy. These measures should allow us to track the evolution of the quantum computers as they are developed and improved, as well as potentially providing further insight into the various sources of error that are present.

Going beyond the reported gate errors, one of the simplest things to measure is the violation of the Mermin inequality~\cite{Mermin1990}, 
\begin{equation}\label{eq: GHZ M}
M_\text{GHZ} = |\la \hat{X}\hat{Y}\hat{Y}\ra + \la \hat{Y}\hat{X}\hat{Y}\ra + \la \hat{Y}\hat{Y}\hat{X} \ra - \la \hat{X}\hat{X}\hat{X} \ra| \leq 2,
\end{equation}
where $\hat{X}, \hat{Y}$ are the $x$ and $y$ Pauli-operators, and we omit the consecutive site labels on the operators. This inequality should hold for a classical theory with local realism. If we consider the GHZ state $|\psi\ra = \frac{1}{\sqrt{2}}(|\!\uparrow\uparrow\uparrow\ra - |\!\downarrow\downarrow\downarrow\ra)$, then $M_\text{GHZ} = 4$ and this bound is maximally violated and can be used to demonstrate the "quantumness" of the machine. In Fig.~\ref{fig: accuracy tests}(a) we show the values computed across 4 consecutive days for the first three qubits used for the $N=6$ simulations. This data shows that we are consistently able to violate the Mermin inequality.

As can be seen in the methods, the fidelities of individual gates do not necessarily reflect the accuracy of a simulation across many qubits. It may therefore make sense to consider more practical measures of quality that more directly relate to the simulations we are performing. We use the physical conservation laws of the evolution to improve the accuracy of our simulations by throwing away measurements of unphysical states. The number of measurements that are kept/thrown away can also be taken as a quantitative measure of the effective accuracy of the machine since for a perfect quantum computer we should find that all measured states are in the physical Hilbert. In Fig.~\ref{fig: accuracy tests}(b) we show the percentage of measured states that are physical for 4 consecutive days, showing the variation in the effective quality of the device.

As an unbiased measure of the quality of our simulation we suggest to compute the identity in the form $\hat{I} = \sqrt{\hat{U}^{-1}(t) \hat{U}(t)}$. By implementing a circuit to perform the forward and backward time evolution, the probability of returning to the initial state measures the accuracy of the implementation of the unitary $\hat{U}(t)$. If the implementation is non-unitary then we should expect decay with an increasing number of Trotter steps. We should also expect decay of this quantity if there are only unitary errors since this quantity takes the form of a Loschmidt echo, which measures the sensitivity of the system to perturbations. It does not rely on any special properties of $\hat{U}(t)$, such as the presence and knowledge of conserved quantities. It therefore provides an unbiased measure of the quality of the implementation of $\hat{U}(t)$ and of the quantum device.

We show the computation of the identity performed on the IBM machine across four consecutive days in Fig.~\ref{fig: accuracy tests}(c). In an ideal machine this quantity should be identically equal to 1 for all times. However, with each additional Trotter step -- corresponding to the observed plateaux -- the accuracy drops significantly. We also note that the behaviour observed in Figs.~\ref{fig: accuracy tests}(b) and (c) are very similar, and both reflect the accuracy of the simulations in the previous sections.

\section*{Discussion}\label{sec: discussion}

Probably the most striking feature of our results from the IBM machines is the low \emph{quantitative} accuracy when compared with exact numerics. Considering the limited system sizes and time scales that we can reach; it highlights the current limitations of these quantum devices. Most importantly, this shows that whilst the number of qubits is now reaching limits beyond the capabilities of classical computers, the error rates and/or the isolation of these quantum computers is not yet sufficient for useful computations, at least for accurately simulating quantum dynamics. Due to the large array of possible error sources, pinning down the most damaging for our simulations is a difficult task, and is an import practical area for future research.

Despite this unfortunate conclusion, we are still able to access a range of \emph{qualitative} physical behaviours demonstrating non-trivial simulations of quantum dynamics. We were able to compute a range of expectation values and two-point correlators, and observe behaviour associated with localization, many-body interactions, and the ballistic spreading of quantum correlations. We also observed the compensation of different energy costs due to on-site potentials and neighbouring site interactions and witnessed the generation of entanglement due to unitary evolution through the QFI. 

The goal of using quantum computers for dynamical simulations is to be able to access systems intractable using classical algorithms. The limitations of the current devices that we have observed in our results demonstrate the need for improved quality of the machines and not simply adding more qubits. In this type of simulation, the number of gates, and thus the real execution time of the quantum circuits, grows linearly with the system size and with the number of Trotter steps. This means that we can estimate that we would need at least an order of magnitude improvement in a combination of the gate fidelities and/or T1/T2 times to get close to achieving this goal.

One of the biggest challenges facing the field of quantum computation is how to deal with errors. Although this is primarily an engineering issue to increase the quality, isolation and control of the devices, there is also the theoretical contribution of error correction methods~\cite{Nielsen2010,PreskillChapt7,Devitt2013}. A particularly promising avenue for error correction is to use surface codes~\cite{Fowler2012,Bravyi2017}. One big advantage of these methods is the moderately low fidelities required for them to work effectively. As the size and quality of the quantum computers increases, it is hoped that these error correction schemes will allow us to rapidly increase their scale, and with it their utility. In the meantime, there may also be more room for practical error \emph{mitigation} schemes, such as the one that we have used, to get the most out of NISQ devices.

While we have considered a range of correlation functions and physical mechanisms, there is still much that can be learnt about the current quantum computers and how well they can simulate quantum dynamics for condensed matter systems. In particular, there are physical mechanisms beyond those that we have considered here. For instance, models with gauge coupling to dynamical gauge fields~\cite{Martinez2016,Klco2018,Wiese2013,Zohar2016}, and the physics of confinement~\cite{Kormos2016,Liu2019}. In these settings there is also hope that interesting physics can be extracted from the short-time dynamics and thus may be suited to the current machines. The combination of disorder and interactions, resulting in the many-body localized phase, is also currently a particularly active area of research~\cite{Abanin2017,Nandkishore2015,Choi2016,Schreiber2015}.
The investigation of the transition between MBL and ergodic dynamics may also benefit in the future from quantum computation. It is notoriously difficult to study numerically due to the requirement of large systems and/or long-time simulations~\cite{Pal2010,Luitz2014,Herviou2018}.

In conclusion, digital quantum simulation is still in its infancy and we have shown that it requires an order of magnitude improvement in fidelity and coherence until it will realistically outperform classical computers, at least applied to dynamical problems of interest in condensed matter physics. However, while it is hard to predict the pace of technological progress, our results will serve as a useful benchmark for improvements in the foreseeable future; and in the long run they will provide a snapshot of capabilities at the beginning of a new quantum simulation era.

\hfill

\begin{center}
	\uppercase{\bf Methods}
\end{center}


\hfill\\\textbf{Implementation: Trotterized Evolution}\\
For our global quench protocol, we first need to prepare the initial state of our system. Since the IBM quantum computers are initialized in the state $|\!\uparrow \uparrow \!\cdots \,\ra$ by default, both of our choices of initial states are tensor product states in the $z$-basis and thus can be created by applications of the Pauli $X$ gate. Next, we need to implement the time evolution, which proceeds by three main steps that we will briefly outline here. 

First, we discretize time, that is, we split the time evolution operator, $\hat{U}(t) = e^{-i\hat{H}t}$, into a sequence of discrete operators, i.e., $\hat{U}(t) = \hat{U}(\Delta t) \cdots \hat{U}(\Delta t)$ with fixed $\Delta t$. Each application of the discrete operator $\hat{U}(\Delta t)$ is called a Trotter step. This is illustrated in Fig.~\ref{fig: trotter}(a) where we apply an increasing number of Trotter steps to reach later times.

In our simulations we fix $\Delta t$ to fix the accuracy of our approximation. However, since we can only implement up to $5$ Trotter steps, we add additional data points by varying $\Delta t$ in the final Trotter step. More explictly, consider Trotter steps $M$ and $M+1$, we can add additional data points by using the evolution operator
\begin{equation}
e^{-i\hat{H}t} \approx \hat{U}(\Delta t)^M \hat{U}(\delta t),
\end{equation}
where $\delta t = \Delta t / r$, where $r$ is the number of data points we want between $t=M\Delta t$ and $t = (M+1)\Delta t$. Since the accuracy of the trotter decomposition $U(\delta t)$ is better than that of $U(\Delta t)$ (see below), these extra data points will have errors intermediate between that of the $M^\text{th}$ and $(M+1)^\text{th}$ steps.

Second, we perform a Trotter decomposition of (trotterize) the unitary evolutions, that is, we approximately decompose the operator $\hat{U}(\Delta t)$ into a sequence of unitaries that act on at most two neighbouring qubits. In the following we use either a basic or symmetric Trotter decomposition, shown in Fig.~\ref{fig: trotter}(b) and (c), respectively. For $m$ Trotter steps of length $\Delta t$, the error of the symmetric decomposition is of order $\mathcal{O}(m(\Delta t)^3)$, compared with $\mathcal{O}(m(\Delta t)^2)$ for the basic decomposition. Note, however, that due to the symmetric structure, when we apply several Trotter steps we can combine several layers of gates. This means that we only need an extra two layers of gates compared with the basic decomposition regardless of the number of Trotter steps. 

Third, we must efficiently decompose these two qubit operators into the gates that can be directly implemented on the quantum device, which are the CNOT gate and arbitrary single qubit unitaries. An efficient decomposition is found in Ref.~\cite{Vatan2004a}, which we summarise below. The result is that if $U\neq0$ then $\hat{B}$ and $\hat{C}$ (defined in the figure caption) can be implemented using three CNOTs and with $U=0$ this can be reduced to only two CNOTs.


\hfill\\\textbf{Trotter Decomposition}\\
In this paper we use a Trotter decomposition (commonly known as a trotterization) of the unitary time evolution operator $\hat{U}(t) = e^{-i\hat{H}t}$. That is, we want to approximate these operators by a sequence of more easily implemented operators, namely those that act on at most two qubits.

As a starting point, consider a Hamiltonian of the form $\hat{H} = \hat{A} + \hat{B}$, where $[ \hat{A},\hat{B}] \neq 0$. Then, since these operators do not commute in general, we have that
\begin{equation}\label{eq: trotter decomp}
e^{-i\hat{H}t} = e^{-i\hat{A}t} e^{-i\hat{B}t} + \mathcal{O}(t^2),
\end{equation}
which can be naturally extended to Hamiltonians that are sums of more than two terms. To use this fact for trotterized evolution we can use the two steps outlined in the main text. We will now go through the details of these steps in more detail, for the Hamiltonian Eq.~\eqref{eq: H}:
\begin{equation}
\hat{H} = - J\sum_{j=1}^{N-1} \left(\hat{\sigma}^x_j\hat{\sigma}^x_{j+1} + \hat{\sigma}^y_j \hat{\sigma}^y_{j+1}\right) + U \sum_{j=1}^{N-1} \hat{\sigma}^z_j \hat{\sigma}^z_{j+1} + \sum_{j=1}^N h_j \hat{\sigma}^z_j,
\end{equation}
which we rewrite here for convenience.

First, we discretize time and split the evolution operator $\hat{U}(t)$ into a product of discrete evolution operators $\hat{U}(\Delta t)$, i.e.,
\begin{equation}\label{eq: discrete time}
e^{-i\hat{H} t} = \left( e^{-i\hat{H} \Delta t} \right)^M,
\end{equation}
where $\Delta t = t/M$ and $M$ is the number of "Trotter steps". Since the Hamiltonian commutes with itself, Eq.~\eqref{eq: discrete time} is exact. To perform time evolution, we will typically fix $\Delta t$ and increase the number of Trotter steps $M$ to reach later times.

The second step is to approximate each of these discrete time operators in a similar manner to Eq.~\eqref{eq: trotter decomp}. For notational simplicity, let us first define the operators
\begin{equation}
\hat{A}_j = e^{-i h_j \hat{\sigma}^z_j \Delta t}, \qquad \hat{B}_j = e^{-i(U\hat{\sigma}^z_j\hat{\sigma}^z_{j+1}-J(\hat{\sigma}^x_j\hat{\sigma}^x_{j+1} +\hat{\sigma}^y_j\hat{\sigma}^y_{j+1}))\Delta t}.
\end{equation}
Using these operators, we can make the approximation
\begin{equation}
e^{-i\hat{H}\Delta t} =  \left( \prod_{j} \hat{A}_j \right) \left( \prod_{j \text{ even}} \hat{B}_j \right) \left( \prod_{j \text{ odd}} \hat{B}_j \right) + \mathcal{O}((\Delta t)^2),
\end{equation}
which corresponds to the schematic quantum circuit show in Fig.~\ref{fig: trotter}(b) in the main text, and which we will refer to as the \emph{basic} trotterization. If we wish to evolve to time $t = M\Delta t$, then we find that the error is $\mathcal{O}(\Delta t)$, which is controlled by the size of the Trotter step $\Delta t$. We can therefore improve the accuracy of the approximation by decreasing $\Delta t$, however, this must be balanced against the cost of needing more Trotter steps, as explained in the main text.


To improve the accuracy of our simulations we can use better approximations to the discrete evolution operators by way of higher-order Trotter decompositions~\cite{Hatano2005}. The leading error term in Eq.~\eqref{eq: trotter decomp} is due to the non-zero commutator $[\hat{A},\hat{B}]$. By compensating for this error, we can increase the order of the leading error term. 

The only higher-order decomposition that we will consider is the symmetrized Trotter step. Let us again start with the simple case of $\hat{H} = \hat{A} + \hat{B}$. The symmetric decomposition would then be
\begin{equation}
e^{-i\hat{H}t} = e^{-i\frac{t}{2} \hat{A}} e^{-i t \hat{B} } e^{-i\frac{ t}{2} \hat{A}}  + \mathcal{O}(t^3).
\end{equation}
The error in this decomposition is of higher-order due to the symmetry which ensures that $U(-t) = U(t)^\dag = U^{-1}(t)$. This means that the even-order error terms vanish and the leading error is $\sim (\Delta t)^3$. See Ref.~\cite{Hatano2005} for more details and for an iterative method for constructing higher-order decompositions.

For the Hamiltonian Eq.~\eqref{eq: H} in question, let us again make some definitions to simplify notation
\begin{equation}
\begin{gathered}
\hat{A}_j = e^{-i h_j \hat{\sigma}^z_j \frac{\Delta t}{2}}, \quad \hat{B}_j = e^{-i(U\hat{\sigma}^z_j\hat{\sigma}^z_{j+1}-J(\hat{\sigma}^x_j\hat{\sigma}^x_{j+1} +\hat{\sigma}^y_j\hat{\sigma}^y_{j+1}))\frac{\Delta t}{2}},\\ \hat{C}_j = e^{-i(U\hat{\sigma}^z_j\hat{\sigma}^z_{j+1}-J(\hat{\sigma}^x_j\hat{\sigma}^x_{j+1} +\hat{\sigma}^y_j\hat{\sigma}^y_{j+1}))\Delta t},
\end{gathered}
\end{equation}
then the symmetric Trotter decomposition is 
\begin{equation}\begin{aligned}
e^{-i\hat{H}\Delta t} =  \left( \prod_{j} \hat{A}_j \right) \left( \prod_{j \text{ even}} \hat{B}_j \right) \left( \prod_{j \text{ odd}} \hat{C}_j \right)\left( \prod_{j \text{ even}} \hat{B}_j \right) \left( \prod_{j} \hat{A}_j \right)\\ + \mathcal{O}((\Delta t)^3) ,
\end{aligned}
\end{equation}
which is shown schematically in Fig.~\ref{fig: trotter}(c) of the main text.

\hfill\\\textbf{Measurement}\\
When making a measurement we will find the system in one of the many-body states $|\lambda\ra$ in this basis, e.g., $|\!\uparrow\downarrow \downarrow \cdots \ra$ or $|\!\downarrow \uparrow \downarrow \cdots \ra$. By performing multiple runs and measurements, we can approximate the probability of measuring each of the many-body states, that is, we can extract $|\alpha_\lambda|^2$, where $\alpha_\lambda$ is the probability amplitude for the state $|\lambda\ra$. These probabilities can then be used to construct the observables. To be more concrete, consider the expectation value of the operator $\hat{\sigma}^z_j$ on site $j$. This is computed as follows,
\begin{equation}
\la \psi(t) | \hat{\sigma}^z_j |\psi(t) \ra = \sum_{\lambda : j = \uparrow} |\alpha_\lambda|^2 - \sum_{\lambda' : j = \downarrow} |\alpha_{\lambda'}|^2,
\end{equation}
where the sums are over all states with the $j^\text{th}$ spin up or down respectively, and $|\alpha_\lambda|^2$ is the proportion of measurements for which we found the state $\lambda$. In all the following experiments we will use 8192 measurements per data point, which means that the statistical error for these local correlators is $\sim 0.01$, which is too small to be included in our figures.


\hfill\\\textbf{Error Mitigation: Physical Hilbert Space}\\
As we noted in the previous section, due to the presence of conserved quantities in the Hamiltonian evolution, the Hilbert space of states splits into those that are physically allowed by the evolution and those that are not. In particular, the models we consider have conserved net magnetization $S_z = \sum_j \hat{\sigma}_j^z$. This fact turns out to be advantageous, and allows us to perform rudimentary error \emph{mitigation} -- a term that we use to distinguish it from scalable error correction methods, since in this case we deal only with the lowest order errors.
The idea is to simply disregard any measurements for states outside of the physical Hilbert space. Let use present a simplified argument for why this might be a good thing to do, where we will first assume that only bit-flips can occur. 

A single bit-flip in the course of the evolution will take us outside of the Hilbert space, and let us denote the probability of this happening as $\Delta$. However, the lowest order of errors within the physical Hilbert space is $\Delta^2$, i.e., we need at least two bit-flip errors to get back to the same total magnetization. Hence, by discarding counts outside of the physical Hilbert space we reduce the leading order error to $\Delta^2$. If the probability, $\Delta$, is sufficiently small, then we can rely on these perturbative arguments. However, if the error rate is large enough, then multiple bit-flip errors can become significant and the error mitigation will be ineffective.

In Fig.~\ref{fig: accuracy tests}(b) we show the percentage of measurements that are within the physical Hilbert space, that is, the percentage of measurements that are kept. For the first data points, we are simply measuring the initial state, which has an average measurement fidelity per qubit of $\sim 95 \%$ leading to a value of approximately $(95\%)^6 \approx 70\%$. At long times the number of retained states stabilises and approaches a value that corresponds to the percentage of states in the total Hilbert space that are physical -- in other words, the percentage probability that a randomly selected state is in the physical subspace. Once this number of discarded measurements is reached, the errors are very large and the error mitigation is no longer effective.

Bit-flips are not the only type of errors that could occur.
As an example, there are also phase errors, which do not necessarily change the net magnetization. However, we note that the constrained data does typically have improved accuracy, and so we use the restriction to the physical Hilbert space for all our subsequent data.


\hfill\\\textbf{Quantum Circuits}\\
Here, we will go over some of the details necessary to implement the trotterized evolution operators on the IBM devices. We will only cover those elements of direct relevance to this paper and refer the reader to Ref.~\cite{Nielsen2010} for an introduction to quantum circuits and quantum computation. We will first introduce the quantum gates that can be implemented on the IBM quantum computers, and then decompose the two qubit unitary operations that appear in our trotterized evolution operator in terms of the elementary one and two qubit gates.

A quantum circuit consists of an array of quantum channels -- which represent the physical qubits -- and a series of quantum gates that are applied to them. These quantum gates are unitary operators that can be applied to one or more of the qubits. The IBM quantum devices can implement the CNOT gate along with an arbitrary single qubit gate, parametrized by three phases. The cobination of these gates forms a universal set that is, any $N$ qubit gate can be implemented using a combination of these gates, and in principle an arbitrary quantum computation can be performed.


There is a collection of single particle gates that are useful for writing quantum circuits. We write down a list of the most frequently used gates and how they are implemented on the IBM machines. Consider the computational basis to be the tensor product of single qubit states in the $z$-basis, i.e., $\{|\!\uparrow\ra, |\!\downarrow\ra\}$. All matrix forms of the gates are given in this basis and all measurements are made in this $z$-basis. It is important to note that gate multiplication reads left to right, whereas matrix multiplication is right to left, i.e. 
$\Qcircuit @C=1em @R=.7em {
	& \gate{B} & \gate{A} & \qw 
} = \Qcircuit @C=1em @R=.7em {
	& \gate{A\cdot B} & \qw 
}$.

Firstly, we have the Pauli matrices, which in the standard quantum information notation are
\begin{equation}
\begin{gathered}
\Qcircuit @C=1em @R=.7em {
	& \gate{X} & \qw 
}	= 
\left(\begin{array}{cc}
0 & 1 \\
1 & 0
\end{array} \right), \qquad
\Qcircuit @C=1em @R=.7em {
	& \gate{Y} & \qw 
}	= 
\left(\begin{array}{cc}
0 & -i \\
i & 0
\end{array} \right),\\
\Qcircuit @C=1em @R=.7em {
	& \gate{Z} & \qw 
}	= 
\left(\begin{array}{cc}
1 & 0 \\
0 & -1
\end{array} \right).
\end{gathered}
\end{equation}
Secondly, we have the Hadamard and the $S$ and $T$ phase gates,
\begin{equation}
\begin{gathered}
\Qcircuit @C=1em @R=.7em {
	& \gate{H} & \qw 
}	= 
\frac{1}{\sqrt{2}}\left(\begin{array}{cc}
1 & 1 \\
1 & -1
\end{array} \right), \qquad
\Qcircuit @C=1em @R=.7em {
	& \gate{S} & \qw 
}	= 
\left(\begin{array}{cc}
1 & 0 \\
0 & i
\end{array} \right),\\
\Qcircuit @C=1em @R=.7em {
	& \gate{T} & \qw 
}	= 
\left(\begin{array}{cc}
1 & 0 \\
0 & e^{i\pi/4}
\end{array} \right).
\end{gathered}
\end{equation}
And finally, we have the $X, Y$ and $Z$ rotation gates,
\begin{equation}
\begin{aligned}
\Qcircuit @C=1em @R=.7em {
	& \gate{R_{x}(\theta)} & \qw 
}	&= e^{-i\frac{\theta}{2}X} =
\left(\begin{array}{cc}
\cos\frac{\theta}{2} & -i\sin\frac{\theta}{2} \\
-i\sin\frac{\theta}{2} & \cos\frac{\theta}{2}
\end{array} \right), \\
\Qcircuit @C=1em @R=.7em {
	& \gate{R_{y}(\theta)} & \qw 
}	&= e^{-i\frac{\theta}{2}Y} =
\left(\begin{array}{cc}
\cos\frac{\theta}{2} & -\sin\frac{\theta}{2} \\
\sin\frac{\theta}{2} & \cos\frac{\theta}{2}
\end{array} \right),\\
\Qcircuit @C=1em @R=.7em {
	& \gate{R_{z}(\theta)} & \qw 
}	&= e^{-i\frac{\theta}{2}Z} =
\left(\begin{array}{cc}
e^{-i\frac{\theta}{2}} & 0 \\
0 & e^{i\frac{\theta}{2}}
\end{array} \right),
\end{aligned}
\end{equation}
which correspond to rotations of the qubit around the $x, y$ and $z$ axes respectively.
All single qubit gates can be written as a product of these rotation gates, up to a phase. This phase is global and is not measurable and can therefore be omitted. In the IBM machine, all single qubit gates can be directly implemented using 
\begin{equation}
\Qcircuit @C=1em @R=.7em {
	& \gate{U_3(\theta,\phi,\lambda)} & \qw 
}	=
\left(\begin{array}{cc}
\cos\frac{\theta}{2} & -e^{i\lambda}\sin\frac{\theta}{2}\\
e^{i\phi}\sin\frac{\theta}{2} & e^{i(\lambda+\phi)}\cos\frac{\theta}{2}
\end{array} \right).
\end{equation}
For slightly less general gates the IBM computer implements either $U_2(\phi,\lambda) = U_3(0,\phi,\lambda)$ or $U_1(\lambda) = U_3(0,0,\lambda)$, which use fewer physical operations and shorter real time. Before running the circuits, we can combine all strings of single qubit gates into a single one of these three single qubit gates, using the functions available in qiskit~\cite{qiskit}.


The most important two qubit gate for our purposes is the CNOT gate
\begin{equation}
\vcenter{\Qcircuit @C=1em @R=2em {
		& \ctrl{1} & \qw \\
		& \targ & \qw
}} \, = \, \left(\begin{array}{cccc}
1 & 0 & 0 & 0 \\
0 & 1 & 0 & 0 \\
0 & 0 & 0 & 1 \\
0 & 0 & 1 & 0
\end{array}\right).
\end{equation}
This gate flips the second qubit depending on the state of the first. This gate allows the two qubits to become entangled, and combined with general single qubit gates forms a set capable of universal quantum computation, see Ref.~\cite{Nielsen2010} for a proof. The CNOT is the only multi-qubit gate currently that can be directly implemented on the IBM quantum machines.

Also of interest to us is the reversed CNOT gate
\begin{equation}
\vcenter{\Qcircuit @C=1em @R=.7em{
		& \targ & \qw &\raisebox{-2.1em}{=} && \gate{H} & \ctrl{1} & \gate{H} & \qw \\
		& \ctrl{-1} & \qw &&& \gate{H} & \targ & \gate{H} & \qw
}} \, = \, \left(\begin{array}{cccc}
1 & 0 & 0 & 0 \\
0 & 0 & 0 & 1 \\
0 & 0 & 1 & 0 \\
0 & 1 & 0 & 0
\end{array}\right),
\end{equation}
where we differentiate the reversed CNOT from the CNOT because of the directionality of the IBM machines, i.e., only CNOTs in a given direction can be implemented along the qubit connections. If a CNOT is applied (programmatically) in the wrong directly, the above transformation using Hadamard gates will be applied by qiskit implicitly. Since the single qubit gate fidelities are an order of magnitude better than that of the CNOTs this transformation is not costly, and these additional gates will often be incorporated into other strings of single qubit gates.


\hfill\\\textbf{Change of Basis}\\
When we make a measurement on the quantum machine it is with respect to a given basis, which we take to be the $z$-basis. However, we may choose to change the basis for several reasons such as: to measure different operators, to prepare an initial state, or to apply a gate which is more efficiently implemented in a different basis.

We will consider only local changes of basis, i.e. a change of basis for the individual qubits. While a general local change of basis can be implemented using the general single qubit gates above, the most frequently used will be those that change from the Z basis to the X or Y basis.

To change to the X basis, we use the Hadamard gate, $H$. This implements the transformation
\begin{equation}
Z \rightarrow X, \qquad X \rightarrow Z, \qquad Y \rightarrow -Y.
\end{equation}
Note that this mapping is its own inverse, which is a result of the Hadamard gate being both unitary and Hermitian.

To change to the Y basis, we use a combination of the Hadamard and $S$ gates. We can implement the basis change with the combination $HSH$, which maps
\begin{equation}
Z \rightarrow  Y, \qquad Y \rightarrow  Z, \qquad X \rightarrow -X.
\end{equation}
Note that this combination of gates is not its own inverse but instead is $HS^\dag H$.


\hfill\\\textbf{Two Qubit Gates}\\
The Trotter decomposition allows us to write the general unitary time evolution operator approximately as a product of single and two qubit unitary operators. We therefore want to find a way to write a general two qubit unitary in terms of the CNOT gate and single qubit gates that can be applied directly on the IBM devices. The optimal decomposition is found in Ref.~\cite{Vatan2004a}. We briefly review the main results of this paper that are of direct relevance to us. 

The optimal decomposition uses the fact that a general matrix in $U(4)$ can be decomposed as $U = (A_1 \otimes A_2) \cdot N(\alpha,\beta,\gamma) \cdot (A_3 \otimes A_4)$~\cite{Kraus2001}, where
\begin{equation}
N(\alpha, \beta, \gamma) = \exp\left[i \left(\alpha\sigma^x \otimes \sigma^x + \beta \sigma^y \otimes\sigma^y + \gamma \sigma^z \otimes\sigma^z\right) \right].
\end{equation}
As a quantum circuit, this can be written as
\begin{equation}
\vcenter{\Qcircuit @C=1em @R=.7em{
		& \multigate{1}{U} & \qw &\raisebox{-2.3em}{=} && \gate{A_3} & \multigate{1}{N(\alpha,\beta,\gamma)} & \gate{A_1} & \qw \\
		& \ghost{U} & \qw &&& \gate{A_4} & \ghost{N(\alpha,\beta,\gamma)} & \gate{A_2} & \qw
}}.
\end{equation}
This operator $N(\alpha,\beta,\gamma)$ is of direct interest to us for quantum dynamics since it is already of the form required for our Trotter decomposition. This gate can be constructed using a minimum of three CNOTs. The optimal decomposition for $N(\alpha,\beta,\gamma)$ is given by the quantum circuit
\begin{equation}
\vcenter{\Qcircuit @C=1em @R=.7em{
		&\qw & \targ & \gate{R_z(\theta)} & \ctrl{1} & \qw & \targ & \gate{R_z(\frac{\pi}{2})} & \qw \\
		& \gate{R_z(-\frac{\pi}{2})} & \ctrl{-1} & \gate{R_y(\phi)} & \targ & \gate{R_y(\lambda)} & \ctrl{-1} & \qw & \qw
}},
\end{equation}
where $\theta = \frac{\pi}{2}-2\gamma, \phi = 2\alpha-\frac{\pi}{2}$, and $\lambda = \frac{\pi}{2}-2\beta$.
Note that in Ref.~\cite{Vatan2004a} they use a different sign convention for the rotation gates. Despite the apparent asymmetry of the decomposition, this sequence of gates is symmetric with respect to swapping the two qubits.


For certain cases of our Hamiltonian, namely when $U=0$, the $N(\alpha,\beta,\gamma)$ gate is more general than we need. By restricting ourselves to $N(\alpha,0,\gamma)$ (plus single qubit basis changes) we can reduce the number of CNOTs required to two. This gives us access to matrices of the form 
\begin{equation}
N(\alpha,0,\gamma) = \exp[i(\alpha \sigma^x\otimes \sigma^x + \gamma \sigma^z\otimes \sigma^z)].
\end{equation}
We can proceed with the help of the so-called Magic Matrix~\cite{Vatan2004a,Kraus2001}
\begin{equation}
\mathcal{M} = \frac{1}{\sqrt{2}} 
\left(\begin{array}{cccc}
1 & i & 0 & 0 \\
0 & 0 & i & 1 \\
0 & 0 & i & -1 \\
1 & -i & 0 & 0
\end{array}\right)
= 
\vcenter{\Qcircuit @C=1em @R=.7em{
		& \gate{S} & \qw & \targ & \qw \\
		& \gate{S} & \gate{H} & \ctrl{-1} & \qw
}}.
\end{equation}
Using this matrix we find $\mathcal{M}^\dag N(\alpha,0,\gamma) \mathcal{M} = e^{i\gamma\sigma^z}\otimes e^{i\alpha\sigma^z}$, which in turn implies that $N(\alpha,0,\gamma) = \mathcal{M} (e^{i\gamma\sigma^z}\otimes e^{i\alpha\sigma^z}) \mathcal{M}^\dag$, since $\mathcal{M}$ is unitary. As a quantum circuit this can be written as
\begin{equation}
\vcenter{\Qcircuit @C=1em @R=.7em{
		&& \raisebox{8pt}{\mbox{$\mathcal{M}^\dag$}} &&&&\raisebox{8pt}{\mbox{$\mathcal{M}$}}& \\
		& \targ & \qw & \gate{S^\dag} & \gate{R_z(-2\gamma)} & \gate{S} & \qw & \targ & \qw \\
		& \ctrl{-1} & \gate{H} & \gate{S^\dag} & \gate{R_z(-2\alpha)} & \gate{S} & \gate{H} & \ctrl{-1} & \qw \gategroup{2}{6}{3}{8}{.7em}{--} \gategroup{2}{2}{3}{4}{.7em}{--} 
}}.
\end{equation}
This gate can be further simplified by noting that a product of single qubit gates is another single qubit gate. Furthermore, $[S,R_z(\theta)] = 0$, and $H R_z(\theta) H = R_x(\theta)$ which gives
\begin{equation}
\vcenter{\Qcircuit @C=1em @R=.7em{
		& \multigate{1}{N(\alpha,0,\gamma)} & \qw &\raisebox{-2.5em}{=} && \ctrl{1} & \gate{R_x(-2\alpha)} & \ctrl{1} & \qw \\
		& \ghost{N(\alpha,\beta,\gamma)} & \qw &&& \targ &  \gate{R_z(-2\gamma)} &  \targ & \qw 
}},
\end{equation}
where we have arbitrarily flipped the circuit with respect to the two qubits.


\hfill\\\textbf{Choosing the Best Qubits}\\
In all our numerics we used between 6 and 10 of the qubits of the IBM machines, which is only a subset of the available 20 qubits. Hence, we wish to find the best such subset so that we get the most accurate results from the machine. We do this by using a simple iterative algorithm, which we will outline here. We note that "best" is a matter of definition involving the balance of many different parameters. We define best to mean the set of qubits that has the lowest average CNOT errors. We do, however, impose restrictions on the allowed measurement error and T2 times.

The algorithm consists of the following steps:
\begin{enumerate}
	\item Restrict the set of allowed qubits to $\tilde{N}$ by discarding any for which the measurement error exceeds a measurement threshold or has a T2 time lower than a given threshold.
	\item List all CNOTs that connect the allowed qubits.
	\item Set value $M = N-1$, where $N$ is the number of qubits.
	\item Construct a restricted list of $M$ CNOTs that have the lowest errors.
	\item From this restricted list of CNOTs, iteratively, construct all chains that connect $N$ qubits.
	\item If no such chain is found and $M<\tilde{N}$, then set $M = M+1$ and repeat from step 4. If no chain is found and $M=\tilde{N}$, then add an extra qubit to the allowed list, increasing $\tilde{N} \rightarrow \tilde{N} + 1$ by increasing the measurement threshold and repeat from step 2.
	\item Once a set of possible chains is found, pick the one with the lowest average CNOT error. 
\end{enumerate}

We note that this algorithm is not necessarily efficient or scalable, but it works for the current size of the quantum computers. See Ref.~\cite{Nishio2019} for an alternative approach. For NISQ devices, we expect that we will similarly want to pick the best subset of qubits, rather than all of them, to achieve greater accuracy simulations and computations. Therefore, an efficient algorithm for large systems is likely to be important.

\begin{figure}[b]
	\centering
	{\hspace*{-10pt}\subfigimg[height=.29\textwidth]{\hspace*{10pt}\textbf{(a)}}{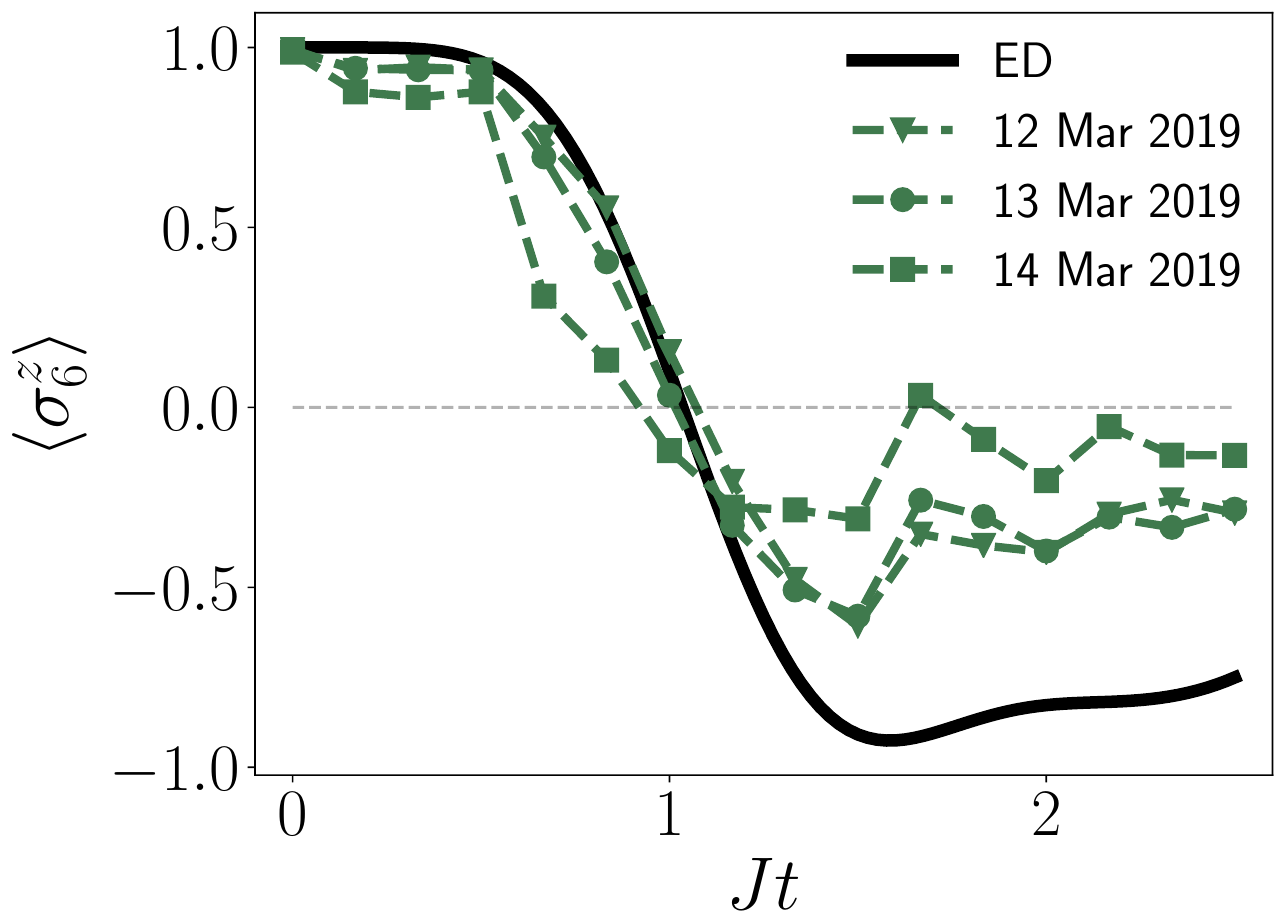}}
	{\hspace*{-10pt}\subfigimg[height=.29\textwidth]{\hspace*{10pt}\textbf{(b)}}{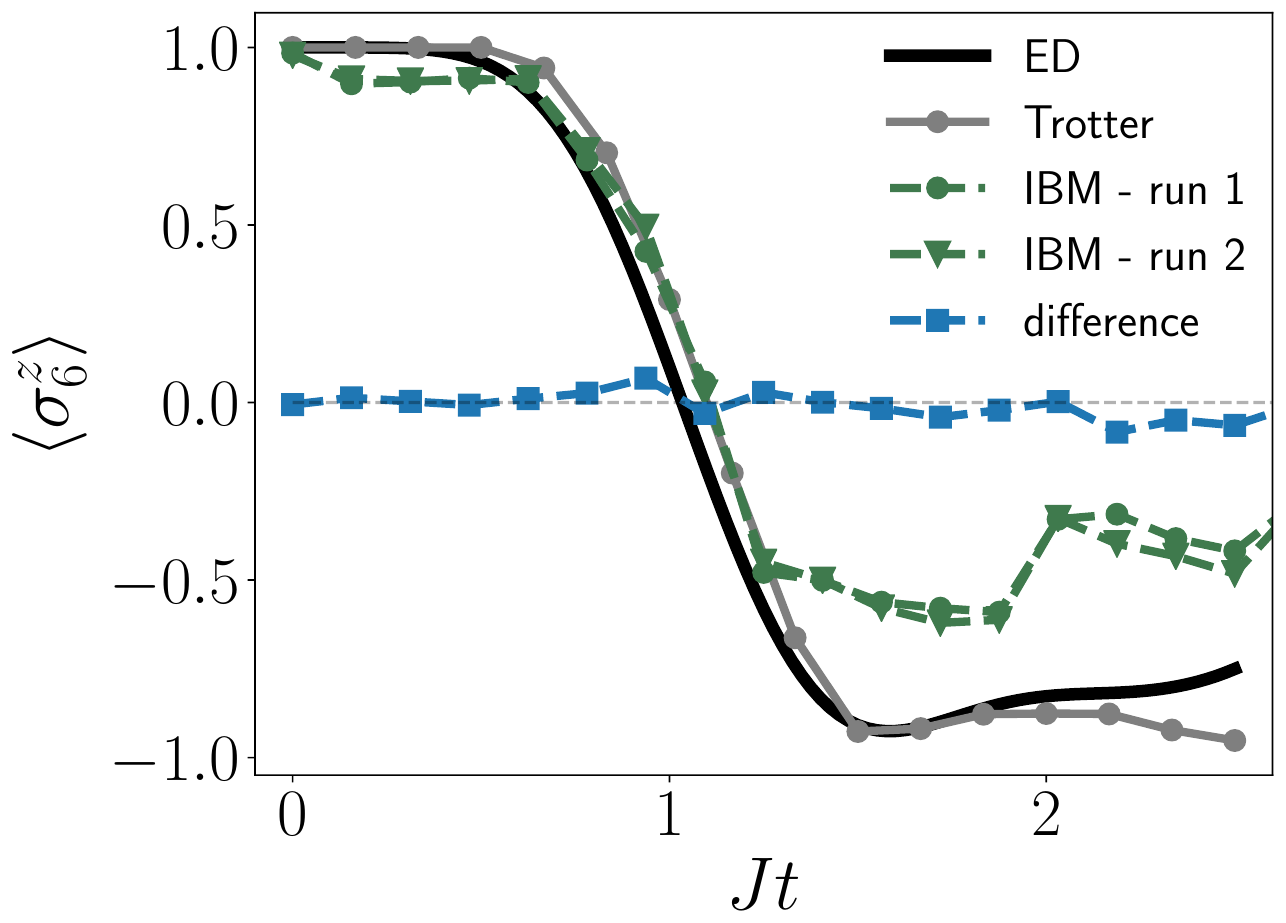}}
	\caption{(a) Simulation of the magnetization on the end spin compared across three consecutive days. (b) Comparison of two separate runs for the local magnetization on the end site after a quench from a domain wall with $N=6$ and Hamiltonian~(\hyperref[case i]{case (i)}). The data sets were obtained on 6 May 2019 with approximately 10 hours between runs. The blue squares show the difference between runs. In both figures we impose the conservation laws, see main text.}\label{fig: local density comparison}
\end{figure}

\begin{figure}[t]
	\centering
	\subfigimg[height=.36\textwidth]{\textbf{\,\qquad(a) 12 Mar 2019}}{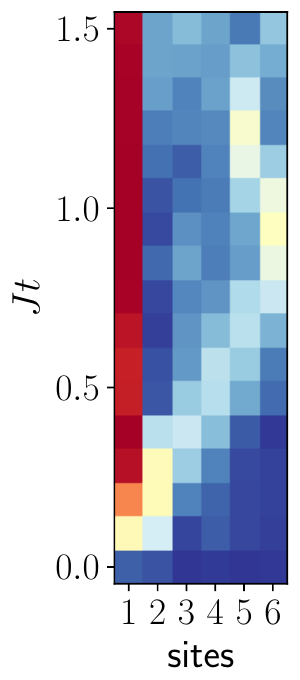}
	\!\!\!\subfigimg[height=.36\textwidth]{\textbf{\!\quad(b) 13 Mar 2019}}{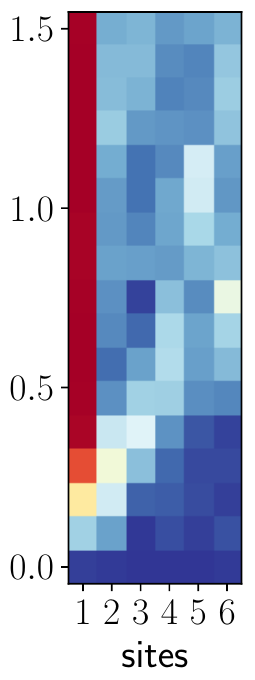}
	\!\!\!\subfigimg[height=.365\textwidth]{\raisebox{-2.5pt}{\textbf{\!\quad(c) 14 Mar 2019}}}{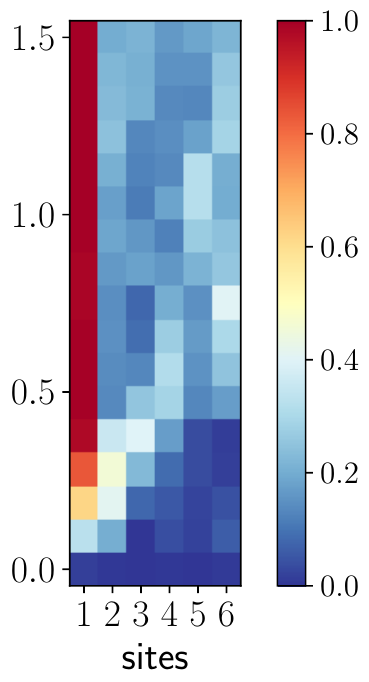}
	\caption{Data computed across three consecutive days for the spin correlator corresponding to Fig.~\ref{fig: correlators} of the main text. The subfigures are labelled by the dates on which the simulation was run.}\label{fig: correlators comparison}
\end{figure}

{
	\rowcolors{2}{white}{gray!15}
	\begin{table*}
		\centering
		\bgroup
		\setlength{\arrayrulewidth}{1pt}
		\setlength{\tabcolsep}{0pt}
		\renewcommand{\arraystretch}{1.2}
		\begin{tabular}{| l c | c | c | c |}
			\hline
			& & 12 Mar 2019 & 13 Mar 2019 & 14 Mar 2019 \\
			\hline
			\;qubits: & 6 \; & [3 2 1 0 5 6] & [3 4 9 14 19 18] & [3 4 9 14 19 18] \\
			& 8 \; & [5 0 1 2 3 4 9 8] & [6 5 0 1 2 3 4 9] & [6 5 0 1 2 3 4 9] \\
			& 10 \; & \;[6 5 0 1 2 3 4 9 14 19]\; & \;[6 5 0 1 2 3 4 9 14 19]\; & \;[6 5 0 1 2 3 4 9 14 13]\; \\
			\hline
			\;readout & min \; & 0.0370 & 0.0290 & 0.0220 \\
			\;error: & avg \; & 0.0442 & 0.0352 & 0.0357 \\
			& max \; & 0.0670 & 0.0420 & 0.0610 \\
			\hline
			\;CNOT & min \; & 0.0127 & 0.0136 & 0.0141 \\
			\;error: & avg \; & 0.0215 & 0.0176 & 0.0192 \\
			& max \; & 0.0288  & 0.0225 & 0.0247 \\
			\hline
			\;T2 times ($\mu$s):\; & min \; & 58.03 & 61.74 & 85.37 \\
			& avg \; & 89.62 & 94.77 & 98.38 \\
			& max \; & 125.01 & 113.54 & 117.71 \\
			\hline
		\end{tabular}
		\egroup
		\caption{Values for IBM machine for the three consecutive days of simulations. The first rows show the optimal qubits as selected by the algorithm in Methods. The remaining rows correspond to the readout errors, CNOT errors, and T2 times for the 6-qubit chain. Data for the 8 or 10 qubit chains, or for all 20 qubits, are not shown.}\label{tab: days}
	\end{table*}
}

\begin{figure*}[t]
	\centering
	\subfigimg[width=.33\textwidth]{\hspace*{10pt}\raisebox{4pt}{\textbf{(a) \qquad\qquad\quad\; 12 Mar 2019}}}{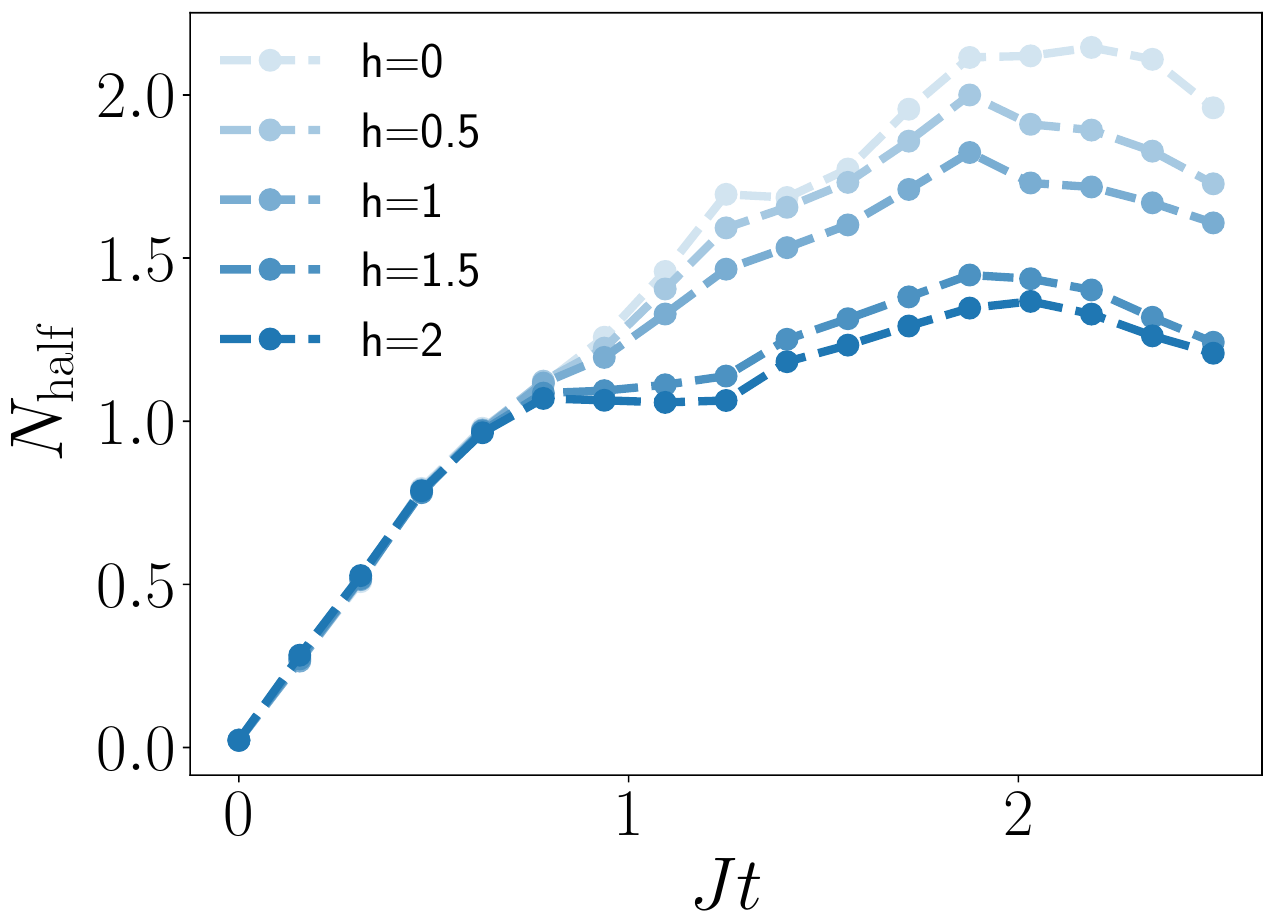}
	\subfigimg[width=.33\textwidth]{\hspace*{10pt}\raisebox{4pt}{\textbf{(b) \qquad\qquad\quad\; 13 Mar 2019}}}{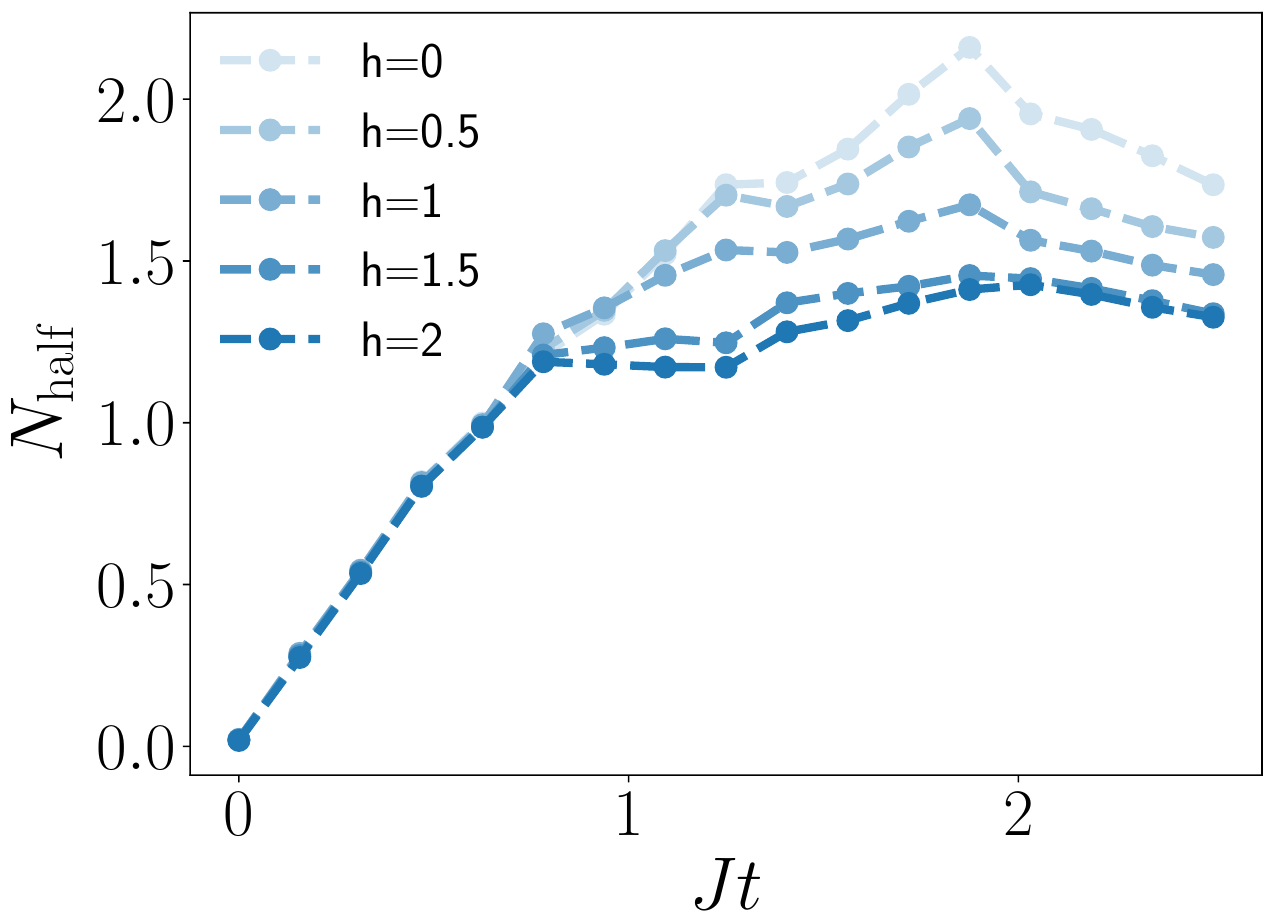}
	\subfigimg[width=.33\textwidth]{\hspace*{10pt}\raisebox{4pt}{\textbf{(c) \qquad\qquad\quad\; 14 Mar 2019}}}{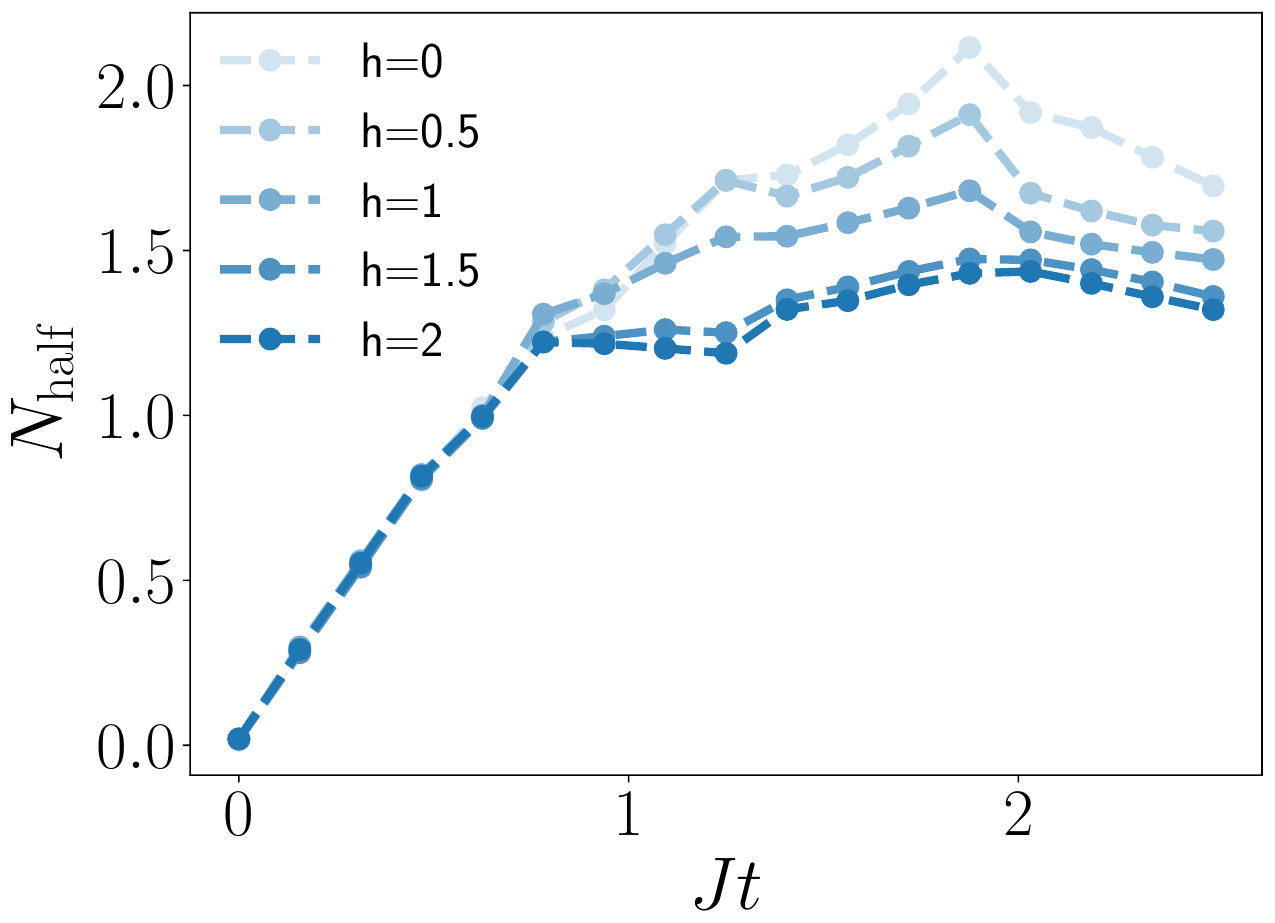}
	\caption{Data computed across three consecutive days, corresponding to Fig.~\ref{fig: disorder}(a). The date for each subfigure is shown at the top.}\label{fig: disorder comparison}
\end{figure*}

\hfill\\\textbf{Data Across Different Days}\\
One practical issue with using the current IBM machines, is the fact that they are recalibrated on approximately a daily basis. Due to the large errors inherent in the devices, this can have a large effect on the results of a simulation and even mean that a different set of qubits is the "best" on different days. This has the unfortunate consequence that the data obtained from the machine will depend on the day on which it was computed. In this section we show a comparison of the results across three days, demonstrating this variability. Thankfully, we also show that the conclusions that we have made in the previous section remain valid independent of the day on which we performed the computations.

First, in Fig.~\ref{fig: local density comparison}(a) we show the constrained IBM data for the local density of the end spin -- corresponding to that shown in Fig.~\ref{fig: tests}(b) -- obtained across the three days. This demonstrates the large fluctuations in our simulations that are due to the different calibrations of the device. For comparison, we show the results from two runs from the same day separated by approximately 10 hours in Fig.~\ref{fig: local density comparison}(a). Unlike in Fig.~\ref{fig: local density comparison}(a) the IBM device was not recalibrated between runs. The difference between the runs (blue squares) is of the order of 5\%, which is comparable to the measurement and statistical errors.

Fortunately, the day-to-day fluctuations do not affect the qualitative behaviour that is simulated by the machine. For example, in Fig.~\ref{fig: correlators comparison} we compare the data for the spin correlator, corresponding to Fig.~\ref{fig: correlators}(c) of the main text. While some of the behaviour is reproduced in all three subfigures -- such as the linear light-cone spreading at short-times -- the quality of the simulations in Fig.~\ref{fig: correlators comparison}(a) is quite clearly better. For instance, we see a much clearer linear spread of the correlations, and it is the only subfigure~\ref{fig: correlators comparison}(a) where the correlations convincingly reach the boundary of the system.

We also consistently see the correct qualitative behaviour for the spread of particles, $N_\text{half}$, in Fig.~\ref{fig: disorder comparison}. Here we show the data corresponding to Fig.~\ref{fig: disorder}(a). While the behaviour differs quantitatively across the days, the qualitative physical behaviour is clear in all plots. This verifies the robustness of the results that we presented above.

Finally, in table~\ref{tab: days} we show the different qubits that were chosen on three consecutive days. We also show the minimum, average and maximum values for the readout errors, CNOT errors and T2 decoherence times for the 6-qubit chain. These values were computed using the calibration data provided by IBM. This table shows that across the three day period, the optimal set of qubits can change and the corresponding values can fluctuate significantly.

While there is clear variation across the three subsequent days that we studied, we point out that the parameters provided by IBM, shown in Table~\ref{tab: days}, are not necessarily good indicators of the quality of the simulation. For instance, the numbers in the table might suggest that on 12 March 2019 the accuracy of the quantum computer was the worst of the three, yet we find the opposite when considering the data for our simulations, see e.g., Fig.~\ref{fig: correlators comparison}. While it is true that the results obtained from the other IBM devices are generally worse, which also reflects a significant difference in gate errors etc., it seems that the small set of numbers in Table~\ref{tab: days} is not able to fully characterize the machine.


\hfill
\begin{acknowledgements}
	We are grateful for discussions with Peter Haynes, Florian Mintert, {\'E}amonn Murray, Abolfazl Bayat, Johnnie Gray, Stefanos Kourtis, Valentin Leeb, and Markus Heyl.
	We acknowledge the Samsung Advanced Institute of Technology Global Research Partnership. A.S. and F.P. were in part funded by the European Research Council (ERC) under the European Union's Horizon 2020 research and innovation programme (grant agreement No. 771537). All data presented is made available in a public GitHub repository~\cite{IBMdata}.
\end{acknowledgements}

\section*{Contributions}

A.S. obtained the data and produced the figures. All authors discussed the results and contributed to the final manuscript.

\iftrue

\end{document}